\documentclass[aps,prc,superscriptaddress,showpacs,nofootinbib,floatfix,twocolumn]{revtex4}

\usepackage{graphicx}   
\usepackage{amsmath}

\def\pt{$p_T$}

\def\gevc{GeV/$c$}
\def\pp{$p+p$}
\def\piz{$\pi^{0}$}
\def\rg{$R_{\gamma}$}

\def\rgmiss{$R_{\gamma}^{\rm miss}$}
\def\rgiso{$R_{\gamma}^{\rm iso}$}

\def\ptt{$p_T^{\rm trig}\,$}
\def\pta{$p_T^{\rm assoc}\,$}
\def\dphi{$\Delta\phi$}

\def\xe{$x_E$}

\def\kt{$k_T$}
\def\jt{$j_T$}
\def\pout{$p_{\rm out}$}
\def\ktrms{$\sqrt{\left<|k_T|^2\right>}$}
\def\jtrms{$\sqrt{\left<|j_T|^2\right>}$}
\def\poutrms{$\sqrt{\left<|p_{\rm out}|^2\right>}$}
\def\pizh{$\pi^0-h$}
\newcommand{\ddecpi}{\mbox{$\delta_{\gamma(\pi)}$}}

\def\gjet{$\gamma$ + jet}
\def\ptpair{$p_T^{\rm pair}$}
\def\veckt{$\vec{k}_T$}
\def\vecjt{$\vec{j}_T$}
\def\vecpout{$\vec{p}_{\rm out}$}
\def\lsim{\raise0.3ex\hbox{$<$\kern-0.75em\raise-1.1ex\hbox{$\sim$}}}
\def\gsim{\raise0.3ex\hbox{$>$\kern-0.75em\raise-1.1ex\hbox{$\sim$}}}

\begin{document}

\title{High $p_T$ Direct Photon and $\pi^{0}$ Triggered 
Azimuthal Jet Correlations and Measurement of $k_T$ for Isolated 
Direct Photons in $p+p$ collisions at $\sqrt{s}$~=~200 GeV}

\newcommand{\abilene}{Abilene Christian University, Abilene, Texas 79699, USA}
\newcommand{\acadsin}{Institute of Physics, Academia Sinica, Taipei 11529, Taiwan}
\newcommand{\banaras}{Department of Physics, Banaras Hindu University, Varanasi 221005, India}
\newcommand{\barc}{Bhabha Atomic Research Centre, Bombay 400 085, India}
\newcommand{\bnlcoll}{Collider-Accelerator Department, Brookhaven National Laboratory, Upton, New York 11973-5000, USA}
\newcommand{\bnlphys}{Physics Department, Brookhaven National Laboratory, Upton, New York 11973-5000, USA}
\newcommand{\caucr}{University of California - Riverside, Riverside, California 92521, USA}
\newcommand{\charlesczech}{Charles University, Ovocn\'{y} trh 5, Praha 1, 116 36, Prague, Czech Republic}
\newcommand{\ciae}{China Institute of Atomic Energy (CIAE), Beijing, People's Republic of China}
\newcommand{\cns}{Center for Nuclear Study, Graduate School of Science, University of Tokyo, 7-3-1 Hongo, Bunkyo, Tokyo 113-0033, Japan}
\newcommand{\colorado}{University of Colorado, Boulder, Colorado 80309, USA}
\newcommand{\columbia}{Columbia University, New York, New York 10027 and Nevis Laboratories, Irvington, New York 10533, USA}
\newcommand{\czechtech}{Czech Technical University, Zikova 4, 166 36 Prague 6, Czech Republic}
\newcommand{\dapnia}{Dapnia, CEA Saclay, F-91191, Gif-sur-Yvette, France}
\newcommand{\debrecen}{Debrecen University, H-4010 Debrecen, Egyetem t{\'e}r 1, Hungary}
\newcommand{\elte}{ELTE, E{\"o}tv{\"o}s Lor{\'a}nd University, H - 1117 Budapest, P{\'a}zm{\'a}ny P. s. 1/A, Hungary}
\newcommand{\fit}{Florida Institute of Technology, Melbourne, Florida 32901, USA}
\newcommand{\fsu}{Florida State University, Tallahassee, Florida 32306, USA}
\newcommand{\gsu}{Georgia State University, Atlanta, Georgia 30303, USA}
\newcommand{\hiroshima}{Hiroshima University, Kagamiyama, Higashi-Hiroshima 739-8526, Japan}
\newcommand{\ihepprot}{IHEP Protvino, State Research Center of Russian Federation, Institute for High Energy Physics, Protvino, 142281, Russia}
\newcommand{\illuiuc}{University of Illinois at Urbana-Champaign, Urbana, Illinois 61801, USA}
\newcommand{\instpasczech}{Institute of Physics, Academy of Sciences of the Czech Republic, Na Slovance 2, 182 21 Prague 8, Czech Republic}
\newcommand{\isu}{Iowa State University, Ames, Iowa 50011, USA}
\newcommand{\jinrdubna}{Joint Institute for Nuclear Research, 141980 Dubna, Moscow Region, Russia}
\newcommand{\kek}{KEK, High Energy Accelerator Research Organization, Tsukuba, Ibaraki 305-0801, Japan}
\newcommand{\kfki}{KFKI Research Institute for Particle and Nuclear Physics of the Hungarian Academy of Sciences (MTA KFKI RMKI), H-1525 Budapest 114, POBox 49, Budapest, Hungary}
\newcommand{\korea}{Korea University, Seoul 136-701, Korea}
\newcommand{\kurchatov}{Russian Research Center ``Kurchatov Institute", Moscow, Russia}
\newcommand{\kyoto}{Kyoto University, Kyoto 606-8502, Japan}
\newcommand{\labllr}{Laboratoire Leprince-Ringuet, Ecole Polytechnique, CNRS-IN2P3, Route de Saclay, F-91128, Palaiseau, France}
\newcommand{\lawllnl}{Lawrence Livermore National Laboratory, Livermore, California 94550, USA}
\newcommand{\losalamos}{Los Alamos National Laboratory, Los Alamos, New Mexico 87545, USA}
\newcommand{\lpc}{LPC, Universit{\'e} Blaise Pascal, CNRS-IN2P3, Clermont-Fd, 63177 Aubiere Cedex, France}
\newcommand{\lund}{Department of Physics, Lund University, Box 118, SE-221 00 Lund, Sweden}
\newcommand{\mass}{Department of Physics, University of Massachusetts, Amherst, Massachusetts 01003-9337, USA }
\newcommand{\muenster}{Institut f\"ur Kernphysik, University of Muenster, D-48149 Muenster, Germany}
\newcommand{\muhlenberg}{Muhlenberg College, Allentown, Pennsylvania 18104-5586, USA}
\newcommand{\myongji}{Myongji University, Yongin, Kyonggido 449-728, Korea}
\newcommand{\nagasaki}{Nagasaki Institute of Applied Science, Nagasaki-shi, Nagasaki 851-0193, Japan}
\newcommand{\newmex}{University of New Mexico, Albuquerque, New Mexico 87131, USA }
\newcommand{\nmsu}{New Mexico State University, Las Cruces, New Mexico 88003, USA}
\newcommand{\ornl}{Oak Ridge National Laboratory, Oak Ridge, Tennessee 37831, USA}
\newcommand{\orsay}{IPN-Orsay, Universite Paris Sud, CNRS-IN2P3, BP1, F-91406, Orsay, France}
\newcommand{\peking}{Peking University, Beijing, People's Republic of China}
\newcommand{\pnpi}{PNPI, Petersburg Nuclear Physics Institute, Gatchina, Leningrad region, 188300, Russia}
\newcommand{\riken}{RIKEN Nishina Center for Accelerator-Based Science, Wako, Saitama 351-0198, JAPAN}
\newcommand{\rikjrbrc}{RIKEN BNL Research Center, Brookhaven National Laboratory, Upton, New York 11973-5000, USA}
\newcommand{\rikkyo}{Physics Department, Rikkyo University, 3-34-1 Nishi-Ikebukuro, Toshima, Tokyo 171-8501, Japan}
\newcommand{\saispbstu}{Saint Petersburg State Polytechnic University, St. Petersburg, Russia}
\newcommand{\saopaulo}{Universidade de S{\~a}o Paulo, Instituto de F\'{\i}sica, Caixa Postal 66318, S{\~a}o Paulo CEP05315-970, Brazil}
\newcommand{\seoulnat}{Seoul National University, Seoul 151-742, Korea}
\newcommand{\stonybrkc}{Chemistry Department, Stony Brook University, Stony Brook, SUNY, New York 11794-3400, USA}
\newcommand{\stonycrkp}{Department of Physics and Astronomy, Stony Brook University, SUNY, Stony Brook, New York 11794, USA}
\newcommand{\subatech}{SUBATECH (Ecole des Mines de Nantes, CNRS-IN2P3, Universit{\'e} de Nantes) BP 20722 - 44307, Nantes, France}
\newcommand{\tenn}{University of Tennessee, Knoxville, Tennessee 37996, USA}
\newcommand{\titech}{Department of Physics, Tokyo Institute of Technology, Oh-okayama, Meguro, Tokyo 152-8551, Japan}
\newcommand{\tsukuba}{Institute of Physics, University of Tsukuba, Tsukuba, Ibaraki 305, Japan}
\newcommand{\vandy}{Vanderbilt University, Nashville, Tennessee 37235, USA}
\newcommand{\waseda}{Waseda University, Advanced Research Institute for Science and Engineering, 17 Kikui-cho, Shinjuku-ku, Tokyo 162-0044, Japan}
\newcommand{\weizmann}{Weizmann Institute, Rehovot 76100, Israel}
\newcommand{\yonsei}{Yonsei University, IPAP, Seoul 120-749, Korea}
\affiliation{\abilene}
\affiliation{\acadsin}
\affiliation{\banaras}
\affiliation{\barc}
\affiliation{\bnlcoll}
\affiliation{\bnlphys}
\affiliation{\caucr}
\affiliation{\charlesczech}
\affiliation{\ciae}
\affiliation{\cns}
\affiliation{\colorado}
\affiliation{\columbia}
\affiliation{\czechtech}
\affiliation{\dapnia}
\affiliation{\debrecen}
\affiliation{\elte}
\affiliation{\fit}
\affiliation{\fsu}
\affiliation{\gsu}
\affiliation{\hiroshima}
\affiliation{\ihepprot}
\affiliation{\illuiuc}
\affiliation{\instpasczech}
\affiliation{\isu}
\affiliation{\jinrdubna}
\affiliation{\kek}
\affiliation{\kfki}
\affiliation{\korea}
\affiliation{\kurchatov}
\affiliation{\kyoto}
\affiliation{\labllr}
\affiliation{\lawllnl}
\affiliation{\losalamos}
\affiliation{\lpc}
\affiliation{\lund}
\affiliation{\mass}
\affiliation{\muenster}
\affiliation{\muhlenberg}
\affiliation{\myongji}
\affiliation{\nagasaki}
\affiliation{\newmex}
\affiliation{\nmsu}
\affiliation{\ornl}
\affiliation{\orsay}
\affiliation{\peking}
\affiliation{\pnpi}
\affiliation{\riken}
\affiliation{\rikjrbrc}
\affiliation{\rikkyo}
\affiliation{\saispbstu}
\affiliation{\saopaulo}
\affiliation{\seoulnat}
\affiliation{\stonybrkc}
\affiliation{\stonycrkp}
\affiliation{\subatech}
\affiliation{\tenn}
\affiliation{\titech}
\affiliation{\tsukuba}
\affiliation{\vandy}
\affiliation{\waseda}
\affiliation{\weizmann}
\affiliation{\yonsei}
\author{A.~Adare} \affiliation{\colorado}
\author{S.~Afanasiev} \affiliation{\jinrdubna}
\author{C.~Aidala} \affiliation{\columbia} \affiliation{\mass}
\author{N.N.~Ajitanand} \affiliation{\stonybrkc}
\author{Y.~Akiba} \affiliation{\riken} \affiliation{\rikjrbrc}
\author{H.~Al-Bataineh} \affiliation{\nmsu}
\author{J.~Alexander} \affiliation{\stonybrkc}
\author{K.~Aoki} \affiliation{\kyoto} \affiliation{\riken}
\author{L.~Aphecetche} \affiliation{\subatech}
\author{R.~Armendariz} \affiliation{\nmsu}
% ...
\collaboration{PHENIX Collaboration} \noaffiliation

\date{\today}

\begin{abstract}

Correlations of charged hadrons of $1 < p_T < 10$ Gev/$c$ with 
high $p_T$ direct photons and $\pi^0$ mesons in the range 
$5 < p_T < 15$ Gev/$c$ are used to study jet fragmentation in the 
$\gamma$+jet and dijet channels, respectively.  The magnitude of 
the partonic transverse momentum, $k_T$, is obtained by comparing to 
a model incorporating a Gaussian $k_T$ smearing.  The sensitivity of 
the associated charged hadron spectra to the underlying 
fragmentation function is tested and the data are compared to 
calculations using recent global fit results.  The shape of the 
direct photon-associated hadron spectrum as well as its charge 
asymmetry are found to be consistent with a sample dominated by 
quark-gluon Compton scattering.  No significant evidence of 
fragmentation photon correlated production is observed within 
experimental uncertainties.

\end{abstract}


\maketitle

\section{Introduction}

\begin{figure*}[thb]
\includegraphics[width=1.0\linewidth]{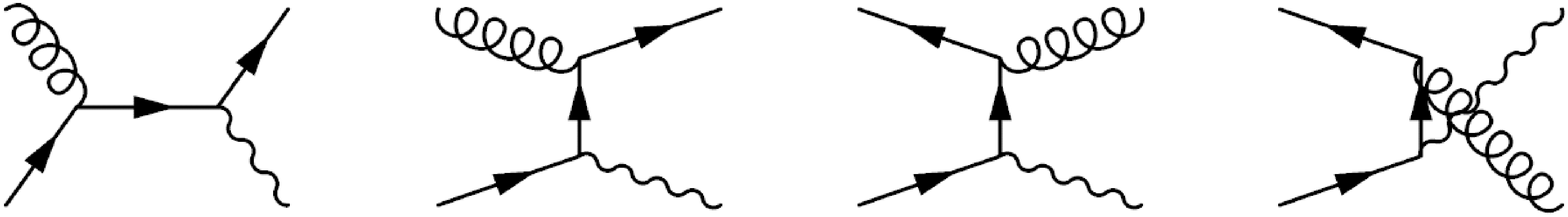}
\caption{Leading order Feynman diagrams contributing to direct 
photon production.  From left to right: $s$ and $u$ channel 
quark-gluon Compton scattering and $t$ and $u$ channel quark 
anti-quark annihilation.}
\label{fig:feyndiag}
\end{figure*}
\begin{figure}[htb]
\includegraphics[width=1.0\linewidth]{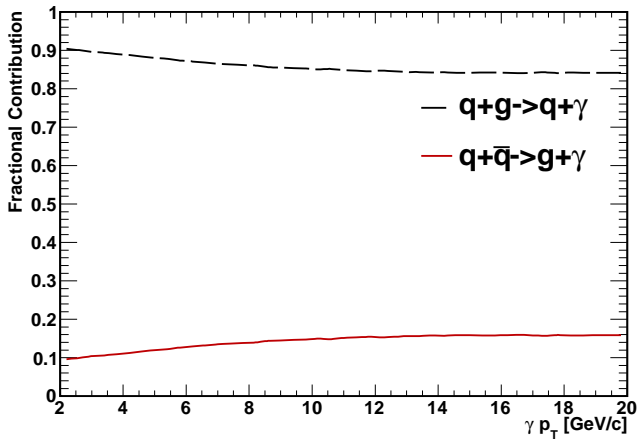}
\includegraphics[width=1.0\linewidth]{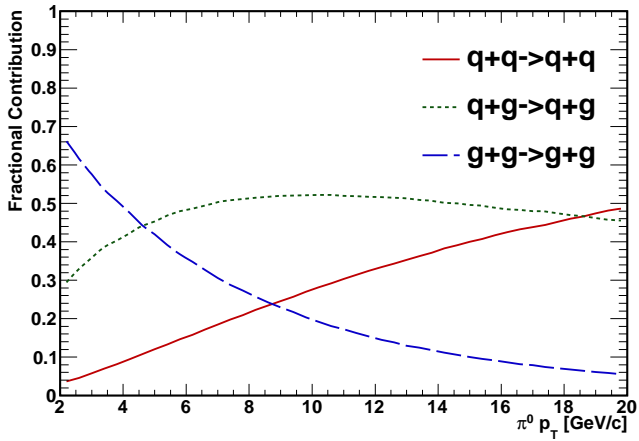}
\caption[]{
The fractional contribution of parton scattering processes to 
(upper) inclusive direct photon and (lower) \piz\ production at LO in \pp\ 
collisions at $\sqrt{s} = 200$ GeV.  The processes $q+\bar{q} 
\rightarrow g+g$ and $g+g \rightarrow q+\bar{q}$ contribute to 
\piz\ triggered events at the level of approximately 1\% and 
0.01\%, respectively, and are not shown.  
\label{fig:processes}}
\end{figure}

Direct photon production in $p+p$ collisions has long been regarded 
as a fundamental observable \cite{ferbel}.  At leading order (LO) 
in perturbative QCD photons are produced directly from the hard 
scattering of partons and are hence independent of non-perturbative 
effects from hadronization.  The LO diagrams, shown in 
Fig.~\ref{fig:feyndiag}, arise from two parton scattering 
processes:  quark-gluon Compton scattering and quark anti-quark 
annihilation.  Figure~\ref{fig:processes}(upper) shows the fractional 
contribution of leading order parton scattering processes for 
$\sqrt{s} = 200$ GeV $p+p$ collisions at midrapidity ($|\eta| < 
0.35$) using the CTEQ6 parton distribution functions~\cite{cteq6}.  
Compton scattering dominates due to the abundance of gluons 
relative to anti-quarks.  Due to the preponderance of gluons in the 
initial state, direct photons have historically been used to probe 
the gluon distribution of the proton~\cite{gluondist1, gluondist2}.  
For comparison, Fig.~\ref{fig:processes}(lower) shows the contributing 
processes to leading order dijet production using a hadron 
trigger, in this case a \piz.  Their relative contributions have a 
non-trivial \pt\ dependence, in large part a consequence of 
triggering on a jet fragment\footnote{The KKP fragmentation 
functions~\cite{kkp} were used in this example.}.  In the final 
state, to the extent that leading order Compton scattering 
dominates, the direct photon is likely to oppose a quark jet.  
Moreover, the transverse momentum of the recoil parton is exactly 
nominally balanced by that of the photon, a feature often exploited 
to determine the energy scale in jet reconstruction~\cite{JES1}.


Beyond LO photons may be produced by bremsstrahlung from a quark.  
Although the rate for hard photon radiation is calculable at NLO, 
photons must be considered part of the jet below an arbitrary 
fragmentation scale.  The contribution of these fragmentation 
photons depends on the non-perturbative parton-to-photon 
fragmentation functions which are poorly constrained relative to 
the fragmentation functions into hadrons~\cite{phofrag1, phofrag2}.  
NLO effects spoil the exact transverse momentum balance between the 
photon and the recoil jet which holds at LO.  Often the 
contribution from fragmentation photons is suppressed by applying 
an isolation criterion to the photon sample~\cite{isoUA1, isoCDF, 
isoD0, isoZEUS, isoH1}.  Typically, the total energy from hadron 
production in a cone around the photon is required to be small 
compared to that of the photon.  The cross section of isolated 
photons may then be compared to NLO calculations with the same 
isolation criterion applied.

At LO, a pair of hard-scattered partons emerges exactly 
back-to-back.  However, due to the finite size of the proton, each 
of the colliding partons has a small transverse momentum on the 
order of 300 MeV~\cite{FFF}.  At NLO, an additional transverse 
momentum component arises from emission of a parton in the initial 
state.  Effects that give rise to a non-zero \ptpair, defined as 
the vector sum of the outgoing parton transverse momenta, are 
collectively referred to as the \kt\ effect, \veckt\ being defined 
as the transverse momentum per parton 
($|k_T | = |p_T^{\ \rm pair}|/\sqrt{2}$).  The component of \kt\ in the 
direction transverse to the outgoing parton pair causes them to be 
acoplanar while the component along the axis of the outgoing parton 
pair imparts to them a momentum imbalance.

For observables such as Drell-Yan production, NLO calculations are 
insufficient to describe the magnitude of the \kt\ effect, since 
multiple soft gluon emission gives an additional contribution 
requiring a resummation of the perturbative 
series~\cite{altarelli}.  Measurements of 
$\langle p_T^{\rm pair}\rangle$ show comparable values for Drell-Yan, 
dijet and diphoton events many of which are compiled in Fig.~30 
of~\cite{ppg029}.

It was argued in~\cite{apanasevich} that NLO pQCD is insufficient 
to describe the world data on direct photon cross sections.  This 
claim was contested in~\cite{aurenche} in which it was argued that 
that inconsistencies amongst the data sets may instead be 
responsible for discrepancies between data and NLO.  Although 
progress has been made, most notably in the form of the 
joint-resummation approach~\cite{laenen}, there is not yet a 
resummed theory which successfully describes all the data.  As a 
consequence, direct photon data are not used in most determinations 
of the parton distribution functions.

Direct photons are also an important observable in heavy-ion 
collisions~\cite{stankus}.  A wealth of evidence indicates that a 
hot, dense state of matter characterized by partonic degrees of 
freedom is produced in these collisions~\cite{whitepaper}.  
Hard-scattered partons are believed to rapidly lose their energy to 
this QCD medium based on the observation of the suppression of 
high-\pt\ hadrons~\cite{ppg003, ppg014}, a phenomenon known as jet 
quenching.  At LO, direct photons play the role of the 
hard-scattered parton (see ~Fig.~\ref{fig:feyndiag}) with the 
distinction that, as color-neutral objects, they do not interact 
strongly with the hot QCD matter.  The absence of any nuclear 
modification for direct photons~\cite{ppg042} therefore acts as a 
control measurement for the jet quenching phenomenon and constrains 
possible contributions from novel sources of induced photon 
production predicted to arise from the interaction of partons with 
the QCD medium~\cite{zakharov, fries}.

Just as photons may be used to determine the jet energy scale in 
\pp\ collisions, they may be used to estimate, on an event-by-event 
basis, the initial energy of the opposite-side parton in heavy-ion 
collisions, an idea first proposed in~\cite{wang}.  The 
distribution of hadrons in the away-side jet reflects the so-called 
medium modified fragmentation function which is the product of the 
jet fragmentation in the dense QCD environment.  Deviations from 
vacuum jet fragmentation, as observed in $p$+$p$ collisions, hence 
should enable tomographic studies of the medium using the energy 
loss of the away-side to probe the density profile of the medium.  
A first, albeit statistics-limited measurement, of direct 
photon-hadron correlations was presented in~\cite{ppg090}.  For a 
recent review of medium-modified fragmentation functions the reader 
is referred to~\cite{mediumFF}.

The interpretation of direct photon triggered correlations in 
heavy-ion collisions necessitates detailed measurements of such 
correlations in $p$+$p$ collisions.  The momentum balance between 
the photon and the opposite-side parton is spoiled due the \kt\ 
effect and by the contribution of fragmentation photons in the 
direct photon sample.  The present work endeavors to study such 
effects.  The remainder of the article is organized as follows.  
In section II we describe elements of the apparatus relevant to the 
measurement of photon-hadron correlations.  In section III the 
methodology of extracting direct photon correlations from the 
background of decay photon-hadron correlations is detailed.  
Section IV presents results on \piz\ and direct photon triggered 
correlations.  Finally, section V interprets the results at the 
partonic level using a simple of LO pQCD calculation coupled with 
phenomenologically-motivated \kt\ smearing.

\section{Detector Description and Particle Identification}

\subsection{The PHENIX Detector}

The PHENIX detector, described in~\cite{overview}, is well suited 
for jet correlation correlations between photons and hadrons.  The 
central arms of the detector are nearly back-to-back in azimuth 
(offset by $22.5^\circ$), each subtending $90^\circ$ and covering 
0.7 units of pseudorapidity around midrapidity.  Each arm contains 
charged particle tracking chambers and electromagnetic 
calorimeters~\cite{centralarm}.

The electromagnetic calorimeters (EMCal) \cite{emcal} consist of two 
types of detectors, six sectors of lead-scintillator (PbSc) 
sampling calorimeters and two of lead-glass (PbGl) \v{C}erenkov 
calorimeters, which measure electromagnetic showers with intrinsic 
resolution $\sigma_E/E = 2.1\% \oplus 8.1\%/\sqrt{E}$ and $0.8\% 
\oplus 5.9\%/\sqrt{E}$, respectively.  The fine segmentation of the 
EMCal ($\Delta\eta \times \Delta\phi \sim 0.01 \times 0.01$ for PbSc 
and $\sim 0.008 \times 0.008$ for PbGl) allows for \piz\ 
reconstruction in the diphoton decay channel out to \pt\ $>$ 20 
\gevc.  In order to measure direct photons over the range $5 < 
p_T <15$ \gevc, \piz\ and $\eta$ must be reconstructed over a 
larger range of $4 < p_T <17$ \gevc\ to account for decay 
feed-down and detector resolution as described below.  Direct 
photon, \piz\, and $\eta$ cross section measurements in \pp\ 
collisions are described in \cite{ppg060, ppg063, ppg055}.  Photon 
candidates of very high purity ($>$ 98\% for energies $> 5$ GeV) 
are selected from EMCal clusters with the use of cluster shower shape 
and charge particle veto cuts.  At large photon \pt\ ($\approx$ 10 
\gevc\ in the PbSc), clusters from \piz\ photon pairs start to 
overlap.  Nearly all of such merged clusters, as well as other 
sources of hadron contamination, have an anomalous shower shape, 
and thus are removed from the analysis.

Charged hadrons are detected with the PHENIX tracking system 
\cite{tracking} which employs a drift chamber in each arm, spanning 
a radial distance of 2.0--2.4 m from the beam axis, and a set of 
pixel pad chambers (PC1) situated directly behind them.  The 
momentum resolution was determined to be $\delta p/p = 0.7\% \oplus 
1.0 \%p$ where $p$ is measured in \gevc.  Secondary tracks from 
decays and conversions are suppressed by matching tracks to hits in 
a second pad chamber (PC3) at a radial distance of $\sim 5.0$ m.  
Track projections to the EMCal plane are used to veto photon 
candidates resulting from charged hadrons that shower in the EMCal.

The data used in this analysis consists of approximately 533 
million photon-triggered events from the 2005 and 2006 $p$+$p$ data 
sets.  The total recorded integrated luminosities during these runs 
were 3.8 (2005) and 10.7 (2006) pb$^{-1}$, respectively.  Events 
were obtained with an EMCal-based photon trigger, described in 
\cite{Adler:2003pb}, which was over 90\% efficient for events with 
a photon or \piz\ in the range of energy used in this analysis.

\subsection{\piz\ and $\eta$ Reconstruction}


\label{sec:pizetarec}

\begin{figure}[htb]
\includegraphics[width=1.0\linewidth]{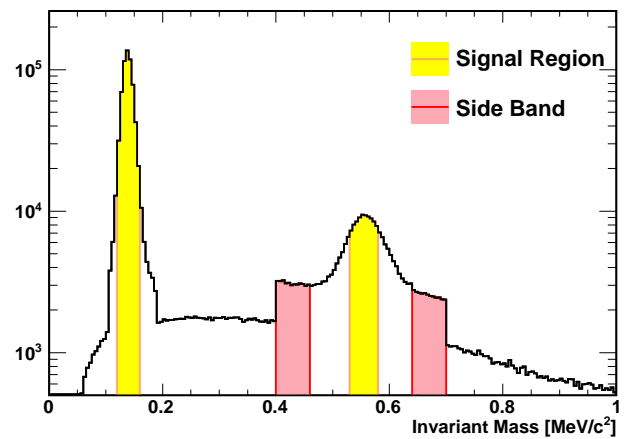}
\caption[]{Di-photon invariant mass distribution for pairs of $5 < 
p_T < 15$ \gevc\ demonstrating \piz\ and $\eta$ reconstruction.  
The signal and side-band regions are indicated.  Offline event 
filtering cuts are responsible for the features at 0.105, 0.185, 
0.4 and 0.7 GeV/$c^2$.  \label{fig:invmass}}
\end{figure}

The background for the present analysis consists of correlated 
decay photon-hadron pairs.  In order to measure this background, 
\piz\ and $\eta$ mesons are reconstructed in the 2-$\gamma$ 
channel, which together are responsible for approximately 95\% of 
the decay photons.  The invariant mass windows for \piz\ and $\eta$ 
mesons are $120$--$160$ and $530$--$580$ MeV/$c^2$, 
respectively, as shown in Fig.~\ref{fig:invmass}.  The rate of 
combinatorial photon pairs is reduced by only considering photons 
of energy $> 1$ GeV.

As discussed in the next section, jet correlations of decay photon 
{\it trigger} particles with charged hadron {\it partner} particles 
are estimated from those of their parent mesons.  The quantity of 
interest is the per-trigger yield, $Y$.  For the \piz, for example, 
the per-trigger yield, $Y_{\pi} $, is the number of \piz-hadron 
pairs divided by the number of \piz\ triggers.  For the given 
photon selection, the effect of combinatorial pairs on the measured 
per-trigger yield was evaluated with PYTHIA to be $<2\%$.  The 
smallness of the effect is due both to the small combinatorial rate 
itself as well as the similarity of correlations from falsely 
matched photons, most of which are themselves from \piz\ decay, to 
the true \piz\ correlations.

In contrast, reconstruction of the $\eta$ meson has a much smaller 
signal-to-background of 1.4--1.6, depending on the \pt\ selection.  
In this case, the per-trigger yield of the combinatorial photon 
pairs is estimated from photon pairs with invariant mass in the 
side-band ranges of 400--460 and 640--700 MeV/$c^2$, 
beyond 3$\sigma$ of the $\eta$ peak.  The procedure for subtracting 
this combinatorial contribution is discussed in~\cite{ppg090} and 
gives rise to a 10\% systematic uncertainty on the $\eta$ 
per-trigger yields.

\section{Subtraction of Decay $\gamma$-Hadron Correlations}
\label{sec:sub}

\subsection{Statistical Subtraction}

A direct photon is defined here to be any photon not from a decay 
process including those from parton to photon fragmentation.  The 
relative contribution of direct and decay photons is expressed in 
terms of the quantity \rg, where \rg\ $\equiv\ $(number of 
inclusive photons)/(number of decay photons).  This quantity has 
been determined in the course of the measurement of the direct 
photon cross section in $\sqrt{s} = 200$ GeV $p$+$p$ collisions 
presented in \cite{ppg060} and its values are given in 
Table~\ref{tab:rgammatable}~\footnote{Note that the values of \rg\ 
are not corrected for efficiency losses due to cluster merging.}.  
In~\cite{ppg090} the yield of charged hadrons per direct photon 
trigger was estimated by a statistical subtraction according to

\begin{equation}
  Y_{\rm direct} = \frac{1}{R_{\gamma} -1} (R_{\gamma}  
  Y_{\rm inclusive} - Y_{\rm decay}),
\label{eq:subtraction}
\end{equation}

\noindent where the trigger particles in the per-trigger yields are 
inclusive photons, decay photons and direct photons as indicated.


The contribution of decay photons to the inclusive photon-hadron 
correlations is determined by weighting the measured \piz\ 
($Y_{\pi}$) and $\eta$ ($Y_{\eta}$) correlations by their 
probability to contribute to a given $Y_{\rm decay}$ bin.  For a 
given \piz\ distribution the number of decay photons from \piz\ in 
the decay photon bin of ($a < p_T ^{\gamma} < b$) is given by

\begin{eqnarray}
N_{\gamma(\pi)}(a < p_T^{\gamma} < b) \nonumber =    \\ 
   \int_a^b
\epsilon_{\gamma}(p_T^{\gamma},p_T^{\pi}) \cdot \mathcal{P}_{\pi}(p_T^{\gamma},p_T^{\pi}) \cdot \epsilon_{\pi}^{-1}(p_T^{\pi}) \cdot N_{\pi}(p_T^{\pi}) \ dp_T ^{\pi}.
\label{eq:decscheme}
\end{eqnarray}

\noindent $\mathcal{P}_{\pi}(p_T^{\gamma},p_T^{\pi})$ is the decay 
probability density for a \piz\ of $p_T^{\pi}$ to decay into a 
photon of $p_T^{\gamma}$.  $\epsilon_{\gamma}$ and $\epsilon_{\pi}$ 
are the single decay photon and \piz\ reconstruction efficiency, 
respectively.  The dependence of $\epsilon_{\gamma}$ on $p_T^{\pi}$ 
is due to cluster merging at high \pt.  For brevity, we define the 
efficiency corrected decay probability density:

   \begin{equation}
   P_{\pi}(p_{T}^{\gamma},p_T^{\pi}) \equiv \epsilon_{\gamma}(p_T^{\gamma},p_T^{\pi}) \cdot \mathcal{P}_{\pi}(p_T^{\gamma},p_T^{\pi}) \cdot \epsilon_{\pi}^{-1}(p_T^{\pi}).
   \end{equation}

\noindent For a finite \piz\ sample the integral in Equation 
\ref{eq:decscheme} becomes a sum and the per-trigger yield of decay 
photons from \piz\ is calculated according to

\begin{equation}
Y_{\gamma(\pi)} = \frac{
\sum_{N_{\pi}} P(p_T^{\gamma},p_T^{\pi}) N_{\pi-h}}
{\sum_{N_{\pi}} P(p_T^{\gamma},p_T^{\pi}) N_{\pi}},
\label{eq:decschemepty}
 \end{equation}

\noindent where the \pt\ limits of the bins have been made implicit 
for brevity.  For a perfect detector, $P$ is calculable 
analytically.  A Monte-Carlo (MC) generator implements the PHENIX 
acceptance and uses Gaussian smearing functions to simulate 
detector resolution according to the known EMCal energy and 
position resolution.  Occupancy effects give rise to an additional 
smearing of the \piz\ and $\eta$ invariant masses.  This effect is 
included in the MC by tuning the resolution parameters to match the 
\piz\ peak widths observed in data.  The uncertainty on the decay 
photon mapping procedure, including the effect of combinatorial 
photon pairs in the \piz\ matching window, was evaluated in PYTHIA 
to be 2\%.  This procedure is described in more detail in 
\cite{ppg090}.

Once the decay photon correlations for \piz\ and $\eta$ triggers 
have been obtained, they are combined according to

\begin{equation}
Y_{\rm decay} = \left( 1/\ddecpi \right) Y_{\gamma(\pi)} + \left(1-1/\ddecpi \right) Y_{\gamma(\eta)}.
\label{eq:etadec}
\end{equation}

\noindent The quantity \ddecpi\ is the ratio of the total number of 
decay photons to the number of decay photons from \piz.  Its value 
was estimated to be $1.24 \pm 0.05$ \cite{ppg060}.  Inefficiencies 
in the detection of photons form \piz\ decay in increase the 
\ddecpi\ slightly, giving a value of 1.28 for decay photons in the 
range 12--15 GeV/c.  Equation \ref{eq:etadec} effectively assigns 
the same per trigger yield to the heavier mesons 
($\omega$,$\eta\prime$,$\phi$,...) as for $\eta$ triggers.  This 
assumption was studied in PYTHIA and found to influence $Y_{\rm 
decay}$ at the level of $ 2\%$.  The total systematic uncertainty 
on the decay photon associated yields contains contributions from 
the $\eta$ sideband subtraction, the value of \ddecpi, the effect 
of hadrons heavier than $\eta$ and the MC decay photon mapping 
procedure,in approximately equal parts.

The per-trigger yields are corrected for the associated charged 
hadron efficiency using a GEANT simulation of PHENIX detector.  The 
quoted yields correspond to a detector with full azimuthal 
acceptance and $|\eta| < 0.35$ coverage.  No correction is applied 
for the $\Delta\eta$ acceptance of pairs.  A \pt-independent 
uncertainty of 8\% was assigned to the charged hadron efficiency.

\subsection{Decay Photon Tagging}

The statistical and systematic uncertainties of the direct photon 
correlation measurement may be improved by event-by-event tagging 
of decay photons.  It is not possible, however, to remove all decay 
photons due the finite acceptance and efficiency for parent meson 
reconstruction.  The remainder must be subtracted according to the 
statistical method described in the preceding section.  Photons are 
tagged when a partner photon of \pt $> $ 500 MeV/$c$ is detected such 
that the pair falls in an invariant mass window of 108--165 or 
500--600 GeV/$c^2$.  In what follows, all such tagged photon pairs 
are presumed to have been correctly identified as coming from the 
same meson parent.  For the $\eta$ meson the rate of combinatorial 
pairs is significant and is subtracted using the side-bands.

Letting the notation {\it tag} denote photons which have been 
tagged as coming from decay sources and the notation {\rm miss} 
denote decay photons which could not be tagged, this residual decay 
component can be statistically subtracted analogously to Equation 
\ref{eq:subtraction} according to

\begin{equation}
Y_{\rm direct} = \frac{1}{R_{\gamma}^{\rm miss}-1}\cdot(R_{\gamma}^{\rm miss}\ Y_{\rm inclusive-tag}-Y_{\rm decay}^{\rm miss}).
\end{equation}

\noindent $Y_{\rm inclusive-tag}$ is simply the per-trigger yield 
of all photons remaining in the sample after tagged decay photons 
are removed.  The effective \rg\ for the sample is

\begin{equation}
R_{\gamma}^{\rm miss} \equiv \frac{N_{\rm inclusive} - N_{\rm tag}}{N_{\rm decay}-N_{\rm tag}} =  \frac{R_{\gamma}}{1-\epsilon^{\rm tag}_{\rm decay}} \frac{N_{\rm inclusive} - N_{\rm tag}}{N_{\rm inclusive}},
\end{equation}

\noindent where the tagging efficiency, $\epsilon^{\rm tag}_{\rm 
decay}$, is $N_{\rm tag}/N_{\rm decay}$.  $\epsilon^{\rm tag}_{\rm 
decay}$ varies from 0.43 in the 5-7 GeV/c bin to 0.53 in the 12-15 
GeV/c bin.

\begin{figure}[htb]
\centering
\includegraphics[width=1.0\linewidth]{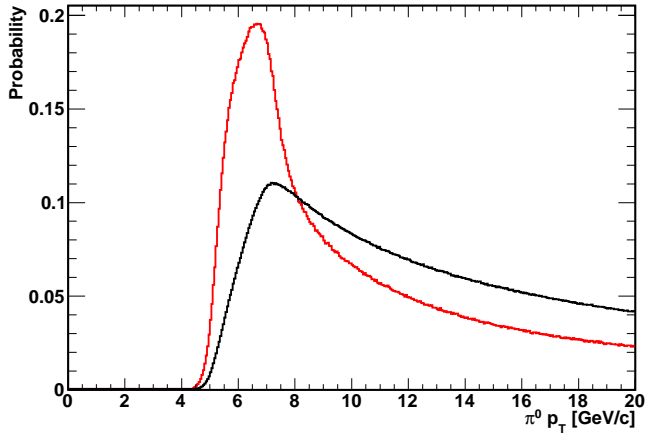}
\caption[Decay probability functions with and without tagging]{The 
decay probability (normalized by area up to $p_T^\pi\ = 20$ \gevc) 
for a \piz\ to decay into a photon with $5 < p_T < 7$ \gevc\ as a 
function of $p_T^{\pi}$ with tagging (red) and without tagging 
(black).} \label{fig:sharkfin}
\end{figure}

In order to determine $Y_{\rm decay}^{\rm miss}$, we define a decay 
photon probability density $P^{\rm miss}$ in which the decay photon 
tagging is performed in the Monte Carlo simulation in the same 
manner as in the data.  Figure~\ref{fig:sharkfin} shows decay 
probability as a function of \piz\ \pt\ for photons in the range $5 
< p_T ^{\gamma} < 7$ \gevc\ ({\it i.e.}, $\int_5^7 P(p_T ^\pi,p_T 
^\gamma) dp_T ^\gamma)$) with and without the decay photon tagging 
applied.  The curves should be interpreted as the probability 
(normalized up to \piz\ \pt\ = 20 GeV/c) for a \piz\ to decay into 
a 5-7 GeV/c photon in PHENIX acceptance as a function of \piz\ \pt.  
The effect of the tagging is most pronounced in this lowest decay 
photon \pt\ bin because the opening angle between the decay photons 
is largest.  Photons which pass the tagging cut are typically 
closer to the parent \piz\ \pt\ than in the case when no tagging 
was applied.

The subtraction procedure defined above assumes that all tagged 
photons were paired correctly.  A systematic uncertainty is 
assigned to account for the possibility that direct photons are 
falsely tagged by accidental combination with a decay photon.  The 
rate of false tagging is estimated from the combinatorial 
background level determined from fits to the invariant mass 
distributions and determined to be as large as 5\% for the highest 
\ptt\ selection.  The direct photon contribution to the falsely 
matched sample is assumed to be the same as in the absence of 
photon tagging (i.e., given by \rg) and the total size of the 
direct photon contribution is taken as the systematic uncertainty.

\subsection{Photon Isolation}

In order to further reduce the decay photon background one may 
impose an isolation requirement, as direct photons are expected to 
be largely produced without much associated hadronic activity.  
Such a requirement has the additional benefit of partially 
suppressing the fragmentation component.  The isolation criterion 
employed in the present analysis is that the sum of the momenta of 
charged tracks and the energy of photon clusters inside a cone of 
radius 0.3 centered around the photon direction is less 10\% of its 
energy.  The cone size is limited by the size of the PHENIX 
aperture which spans 0.7 units of pseudorapidity.  For photons near 
the edge of the detector the isolation cone lies partially outside 
of the PHENIX acceptance.  In order to reduce the impact of the 
acceptance on the photon isolation, fiducial cuts are applied such 
that photons are required to be ~$\sim 0.1$ radians from the edge 
of the detector in both $\eta$ and $\phi$.

As was the case for decay photon tagging, a residual decay 
component must be statistically subtracted after the isolation cuts 
have been applied.  The per-trigger yield of isolated direct photon 
is obtained according to

\begin{equation}
Y_{\rm direct}^{\rm iso} = \frac{1}{R_{\gamma}^{\rm iso}-1} (R_{\gamma}^{\rm iso}\ Y_{\rm inclusive-tag}^{\rm iso}-Y_{\rm decay}^{\rm iso}) 
\end{equation}

Note that the label {\it iso} ({\it niso}) denotes photons which 
are both isolated (non-isolated) and were not removed by the 
tagging cuts.  The value of \rg\ corresponding to the photon sample 
after both types of event-by-event cuts is

\begin{eqnarray}
R_{\gamma}^{\rm iso} 
\equiv \frac{N_{\rm inclusive}-N^{\rm tag}_{\rm decay}-N^{\rm niso}_{\rm inclusive}}{N^{\rm decay} - N^{\rm tag}_{\rm decay} - N^{\rm iso}_{\rm decay}} \nonumber \\
= \frac{R_{\gamma}}{(1-\epsilon^{\rm tag}_{\rm decay})(1-\epsilon^{\rm niso}_{\rm decay})} \frac{N_{\rm inclusive} - N_{\rm tag} - N_{\rm isolated}}{N_{\rm inclusive}}.
\label{eqn:isorg}
\end{eqnarray}

The efficiency with which the isolation cut removes decay photons, 
$\epsilon^{\rm niso}_{\rm decay}$, is not known {\it a priori}, 
since an unknown fraction of the non-isolated photons are direct.  
In order to evaluate the isolation efficiency we apply the 
isolation cut at the level of the parent mesons and use the decay 
probability functions to map the effect to the daughter photon \pt.  
For the example of the \piz,
 
\begin{eqnarray}
\epsilon^{\rm niso}_{\gamma(\pi)} \equiv
\frac{N^{\rm niso}_{\gamma(\pi)}(p_T^{\gamma})}{N_{\gamma(\pi)}^{\rm iso}(p_T^{\gamma}) + N_{\gamma(\pi)}^{\rm niso}(p_T^{\gamma})} \nonumber \\
= \left(1 +
\frac{\sum_{\pi} P^{\rm miss}_{\pi}(p_T^{\pi},p_T^{\gamma}) \cdot N^{\rm iso}_{\pi}(p_T^{\pi})}{\sum_{\pi} P^{\rm miss}_{\pi}(p_T^{\pi},p_T^{\gamma}) \cdot N^{\rm niso}_{\pi}(p_T^{\pi})} \right)^{-1}.
\end{eqnarray}

\noindent We have implicitly exploited the fact that the tagging 
probability is independent of the isolation requirement.  
$\epsilon^{\rm niso}_{\gamma(\pi)}$ varies from 0.4 in the 5-7 
GeV/c bin to 0.48 in the 12-15 GeV/c bin.

The per-trigger yield of isolated (and tagged) \piz\ decay photons 
can be calculated according to

\begin{equation}
Y_{\gamma(\pi)}^{\rm iso} = \frac{
\sum_{N_{\pi}} P^{\rm miss}_{\pi}(p_T^{\gamma},p_T^{\pi}) N_{\pi-h}^{\rm iso}}
{\sum_{N_{\pi}} P^{\rm miss}_{\pi}(p_T^{\gamma},p_T^{\pi}) N_{\pi}^{\rm iso}}.
 \label{eq:tagschemepty}
\end{equation}

\begin{figure}[htb] 
\includegraphics[width=1.0\linewidth]{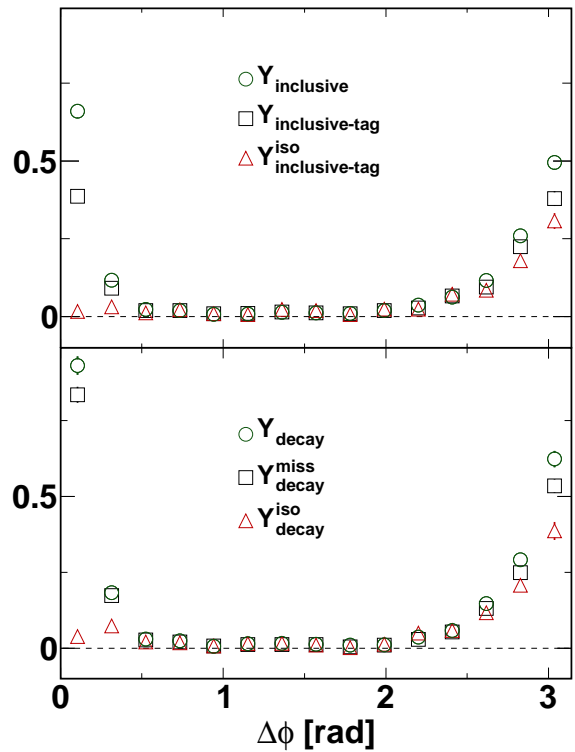} 
\caption[]{ Per-trigger yields for inclusive (top) and decay 
photons (bottom) with no event-by-event cuts, with decay photon 
tagging and with decay photon tagging and the isolation cut.  The 
systematic uncertainties on the decay photon triggered yields are 
not shown.
\label{fig:dphi_inc_dec}}
\end{figure}

\begin{figure*}[htb]
\includegraphics[width=1.0\linewidth]{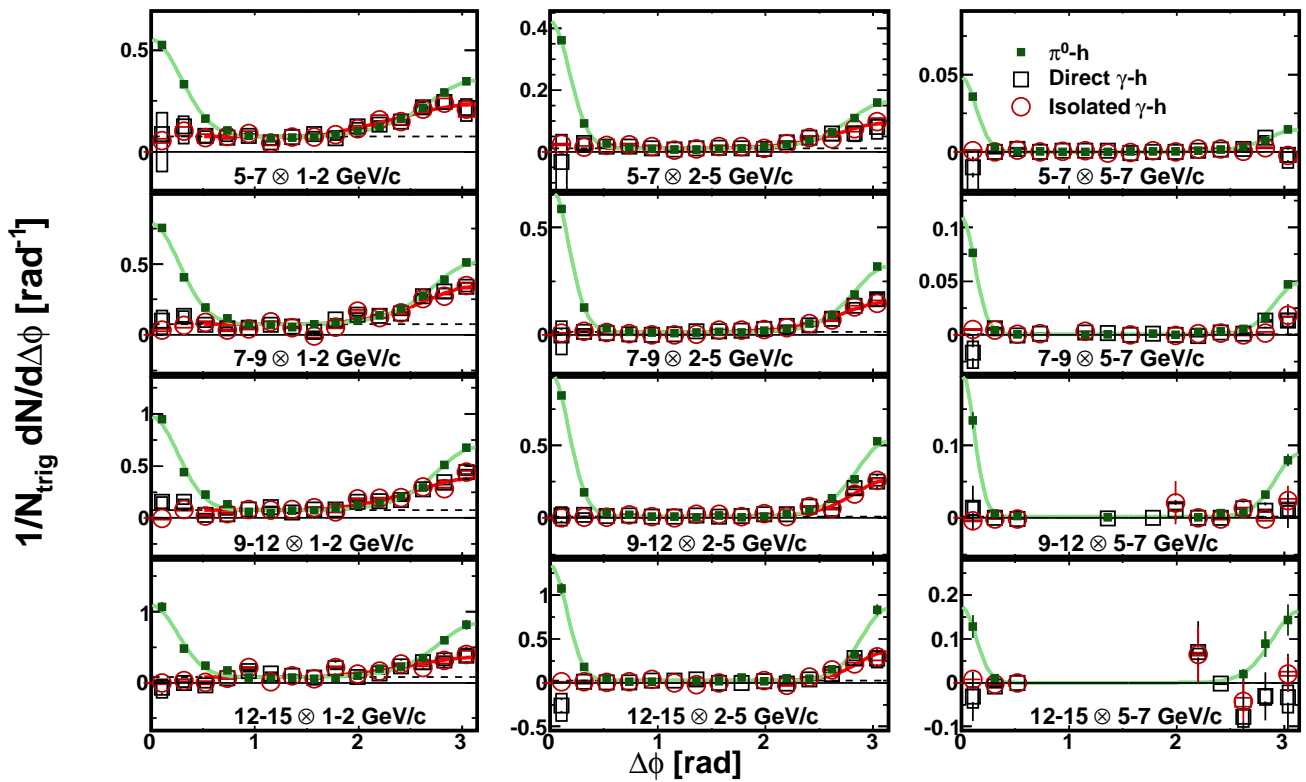} 
\caption[]{Yield per trigger of charged hadrons associated with 
\piz, direct and isolated direct photons triggers as a function of 
\dphi.  The lines correspond to fits which are described in Section 
\ref{sec:pout}.  The constant level, which estimates the 
contribution from the underlying event, is shown as a dashed line 
(only visible in the lower \pta\ bins).  \label{fig:dphi}}
\end{figure*}

\noindent Azimuthal correlations for inclusive and decay photons 
for $9 < $ \ptt $ < 12$ and $2 < $\pta$ < 5$ \gevc\ are shown in 
Fig.~\ref{fig:dphi_inc_dec}.  Also shown are the correlations after 
the tagging and isolation+tagging cuts have been applied.  The 
tagging causes a reduction in the near-side yield of inclusive 
photons confirming that direct production is subsequently enhanced.  
The decay photon associated yield is also smaller after tagging, 
due to the fact that the remaining photons have a smaller mean \pt, 
as can be ascertained from the decay probability function shown in 
Fig.~\ref{fig:sharkfin}.  The isolation cut causes a further 
reduction in the yields, driven by the fact that the fraction of 
the jet momentum carried by the trigger is larger for isolated 
photons.  Table~\ref{tab:rgammatable} shows the \rg\ and its 
effective values for the tagged and tagged+isolated samples.

\begin{table}[t]
\caption{\rg\ and its effective values 
for the photon sample with tagging implemented ($R_{\gamma}^{\rm miss}$)
and the sample with tagging and isolation implemented 
($R_{\gamma}^{\rm iso}$).
}
\label{tab:rgammatable}
\begin{ruledtabular}  \begin{tabular}{cccc}
     \pt\ [\gevc] & \rg\ & \rgmiss\ & \rgiso\   \\ \hline
    5-7 & 1.19 $\pm$ 0.06 & 1.32 $\pm$ 0.11 & 1.38 $\pm$ 0.12  \\
    7-9 & 1.33 $\pm$ 0.05 & 1.67 $\pm$ 0.13 & 1.92 $\pm$ 01.4  \\
    9-12 & 1.54 $\pm$ 0.05 & 2.22 $\pm$ 0.18 & 2.87 $\pm$ 0.22   \\
    12-15 & 1.80 $\pm$ 0.11 & 2.69 $\pm$ 0.36 & 4.02 $\pm$ 0.50  \\
  \end{tabular} \end{ruledtabular}
\end{table}

\section{Results}

\subsection{Azimuthal Correlations}

Essential features of dijet or $\gamma + $jet production are 
evident in azimuthal two particle correlations.  
Figure~\ref{fig:dphi} shows the per-trigger yields of associated 
charged hadrons as function of \dphi\ for \piz, direct photon 
(using decay photon tagging) and isolated direct photon triggers.  
The strong near-side correlation for \piz\ triggers is largely 
absent for direct photon triggers as expected from a sample 
dominated by photons produced directly in the hard scattering.  On 
the away-side the direct photon triggered yields are generally 
smaller than for \piz\ triggers.  Note that the error bars on the 
direct photon yields, which are from \rg\ and from the decay photon 
yields, vary bin-to-bin depending on the relative values of $Y_{\rm 
inclusive}$ and $Y_{\rm decay}$.  The 8\% normalization uncertainty 
on the charged hadron yields is not shown on this plot and all 
those which follow.

\subsection{Near-side Correlations}

\begin{figure}[htb]
\includegraphics[width=1.0\linewidth]{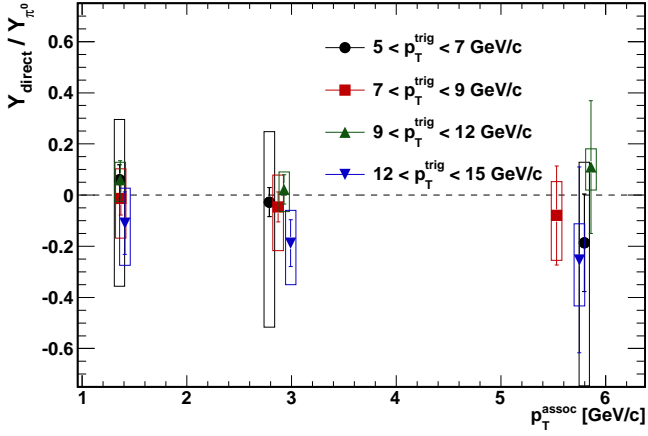} 
\caption[]{The ratio of the direct photon to \piz\ near-side ($\phi 
< \pi/2$) associated yields as a function of \pta.  The points are 
placed according to $< p_T^{\rm assoc}>$ for the isolated direct 
photon triggered sample.  \label{fig:NS}}
\end{figure}

The use of \gjet\ events, whether to calibrate the energy of the 
away-side jet in \pp\ collisions or to study its energy loss in 
nuclear collisions, relies on a detailed understanding of photons 
produced in jet fragmentation, which may be considered a background 
in this regard.  Evidence of fragmentation photons may be observed 
indirectly via near-side correlations of direct photon-hadron 
pairs.  Figure~\ref{fig:NS} shows the ratio of the near-side yield 
of direct photon triggers to that of \piz\ triggers for various 
\pt\ selections.  The ratio is generally consistent with zero, 
albeit within fairly larger uncertainties.  Such larger 
uncertainties on the near-side direct photon per-trigger yields are 
due to the large near-side associated yields for inclusive and 
decay photon triggers compared to the small difference between 
them.

The ratio $Y_{\rm direct}/Y_{\pi^0}$ is not a direct measurement of 
the relative contribution of fragmentation photons to the direct 
photon sample.  This quantity is sensitive to both the overall 
contribution from fragmentation photons as well as the difference 
between parton-to-photon and parton-to-\piz\ fragmentation 
functions and possibly a different number of quark vs.  gluon jets 
in the \piz\ triggered and photon triggered samples.  As an 
illustrative example, if one assumes that the photon and \piz\ 
fragmentation functions are similar in shape, it is reasonable to 
fit the ratio of the per-trigger yields to a constant value as a 
function of $p_T^{\rm assoc}$.  The results of such fits, shown in 
Table \ref{tab:NS} show that above 7 \gevc\ the near-side yield 
associated with direct photons is constrained to be smaller than 
15\% of the \piz\ associated yields.  It should be emphasized that 
although such an observation is compatible with a small yield of 
fragmentation photons, it may also reflect a different shapes of 
the fragmentation function into photons compared to the 
fragmentation function into neutral pions.

\begin{table}[htb]
\caption{Ratio of constant fits to the near-side yields as a 
function of \pta\ using isolated direct photon triggers to the same 
quantity using \piz\ triggers.}
\begin{ruledtabular}   \begin{tabular}{cccc}
     \ptt\ [\gevc] & Ratio from fit & Stat.  & Sys.    \\ \hline
    5-7 & 0.01 & 0.04 & +0.26 - 0.46  \\
    7-9 &  -0.03 & 0.04 & +0.12 - 0.16   \\
    9-12 & 0.04  & 0.04 & +0.07 - 0.09     \\
    12-15 &  -0.16 & 0.07 & +0.14 - 0.17    \\
  \end{tabular}  \end{ruledtabular} 
\label{tab:NS}
\end{table}

\subsection{Hard Scattering Kinematics using  \xe\ and \pout}
\label{sec:jetkinematics} 

Due to the effects of hadronization, we do not have direct access 
to the parton kinematics and therefore can measure neither the 
fragmentation functions nor the magnitude of the \kt\ effect 
directly.  However, to the extent that the LO Compton scattering 
process (see Fig.~\ref{fig:feyndiag}) dominates, direct photons may 
be considered to play the role of the hard scattered parton.  For 
both the case of isolated photon and \piz\ triggered correlations 
we construct a simple model to extract the parton-level kinematics 
from the data, as described in the next section.  To facilitate the 
interpretation of the data we choose a set of observables which are 
appropriately sensitive to the quantities of interest.

The fragmentation function is not directly measurable via two 
particle correlations since the jet momentum is not determined.  
If the near-side jet \pt\ were fixed by that of the trigger 
particle and the jets were balanced, one could measure the 
fragmentation function of the away-side jet by measuring the 
associated yield as a function of the partner \pt.  For the case of 
dihadron correlations, however, the $Q^2$ of the hard scattering 
varies with both the trigger and partner \pt\ selection, which are 
both jet fragments \cite{ppg029}.  Direct photon triggers, on the 
other hand, should provide a nearly mono-energetic sample of jets 
for fixed photon \pt.  The quantity $\hat{x}_h$ measures the 
transverse momentum imbalance between the trigger and associated 
parton:

\begin{equation}
\hat{x}_h \equiv \hat{p}_T^{\rm assoc}/\hat{p}_T^{\rm trig},
\end{equation}

\noindent where the hat symbol denotes partonic level quantities.  
At LO $\hat{x}_h = 1$, but may deviate from unity due to the 
\kt\ effect.  Given a falling $\hat{p}_T $ spectrum, $\hat{x}_{\rm 
h}$ is typically less than one due to the trigger bias.
 
The quantity \xe\ measures the \pt\ balance between the trigger and 
associated particles \cite{Darriulat:1976eh}.  It is defined in 
analogy to the fragmentation function variable $z$ by substituting 
the the \pt\ of the trigger particle for that of the jet and taking 
the projection of the associated particle onto the trigger axis in 
the azimuthal plane:

\begin{equation}
x_E \equiv -\frac{\vec{p}_T^{\ \rm trig} \cdot \vec{p}_T^{\ \rm assoc}}{|p_T^{ \rm trig}|^2} = - \frac{|p_T^{\rm assoc}|}{|p_T^{\rm trig}|} \cos{\Delta \phi}.  \label{eq:xEdef}
\end{equation}

\noindent For \ptt $\approx \hat{p}_T ^{\rm trig}$ (which is not 
the case for jet fragments measured at fixed $p_T^{\rm trig}$), the 
\xe\ distribution approximates the fragmentation function, $D(z)$.  
Hence, the \xe\ distribution for direct photon triggers should 
scale with \ptt, in the same way as the fragmentation functions 
approximately scale with $Q^2$.

\begin{figure}[htb]
\centering
\includegraphics[width=1.0\linewidth]{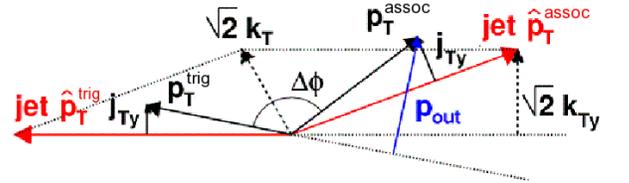}
\caption[A diagram showing the kinematics underlying the 
measurement of jet correlations between back-to-back particles]{ 
\label{fig:jets} A diagram showing the kinematics underlying the 
measurement of jet correlations between back-to-back particles, 
adapted from \protect\cite{ppg029}.
}
\label{fig:ppg029diagram}
\end{figure}

Beyond LO pQCD the outgoing parton pair may acquire a net \pt\ as 
depicted in Fig~\ref{fig:ppg029diagram}.  The pair may acquire a 
net momentum along the both the dijet axis and orthogonal to it.  At 
the hadron level, the kinematics along the jet axis are dominated 
by the effect of jet fragmentation.  The \kt\ effect is therefore 
best observed by measuring the momentum of jet fragments in the 
direction orthogonal to the parton pair.
 Again using the trigger particle direction as a proxy for that of the parton, we define \vecpout\ as a vector transverse to $p_T^{\ \rm trig}$ of magnitude

\begin{equation}
|p_{\rm out}| = |p_T^{\rm assoc}| \sin{\Delta \phi}.
\end{equation}

\noindent \pout\ also contains a contribution from $j_T $ which is 
the momentum transverse to the jet axis imparted to the jet 
fragment in the course of the parton showering process.  Note that 
the presence of $|p_T^{\rm assoc}|$ in \pout\ implies a dependence 
on the fragmentation function of the away-side jet.  Similarly the 
longitudinal (along the dijet axis) component of \veckt\ can play a 
role in the \xe\ distributions.  Such a mixing of longitudinal and 
transverse effects is an unavoidable consequence of hadronization.


\subsection{ \xe\ Distributions}

\begin{figure}[ht]
\includegraphics[width=1.0\linewidth]{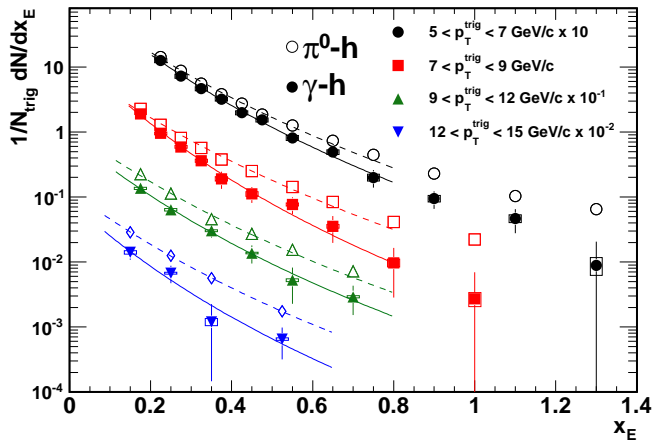}
\caption[\xe distributions]{\label{fig:xedists} Away-side charged 
hadron yield per \piz\ trigger (open symbols) and isolated direct 
photon trigger (closed symbols) as a function of \xe, which is 
equivalent to $z_T$ in the collinear limit $cos(\Delta \phi)$ = 1. 
The lines 
represent fits to the modified Hagedorn function described in the 
text.}
\end{figure}

Figure \ref{fig:xedists} shows the away-side (\dphi\ $> \pi/2$) 
charged hadron yield per trigger as a function of \xe\ for both 
\piz\ and isolated direct photon triggers.  The isolated direct 
photon triggered data are obtained by the same procedure outlined 
in Sec.~\ref{sec:sub} that was used to obtain the \dphi\ 
distributions presented in the previous section.  The isolated 
direct photon \xe\ distributions are significantly steeper than 
those of \piz.  This is to be expected because, modulo the \kt\ 
smearing, the the \xe\ distribution opposite to an isolated direct 
photon should be the fragmentation function of the outgoing parton, 
predominantly the quark from the Compton diagram 
(Fig.~\ref{fig:feyndiag}), with a small gluon admixture from the 
annihilation diagram.

The \xe\ distributions have been fit using a modified Hagedorn 
function defined by

\begin{equation}
\frac{dN} {dx_E} \approx {N (n-1)}
\frac{1} {\hat{x}_h} 
\frac{1} {(1+ \frac{x_E}{\hat{x}_h} )^n},  
\label{eq:mHag}
\end{equation}

\noindent where $n$ is the power-law dependence of the inclusive 
invariant \piz\ \pt\ spectrum.  This functional form was shown to 
describe the \piz\ triggered \xe\ distribution given the following 
simplifying assumptions:  The hadron is assumed to be collinear 
with the parton direction, i.e., $\cos{\Delta \phi}\approx 1$ in 
Eq.~\ref{eq:xEdef}, the underlying fragmentation functions ($D(z)$) 
are assumed to take an exponential form and $\hat{x}_h$ is 
taken to be constant as a function of \xe.  The fit was found to 
perform reasonably well for the range below \xe\ $\lsim\ 0.8$.  
The results of the fits in this range are shown in 
Table~\ref{tab:xepi0}.  The $\chi^2$ values are rather large, 
particularly at low \piz\ \pt, indicating that the data are not 
perfectly described by the modified Hagedorn function, which is 
perhaps not surprising given the good statistical precision of the 
\piz\ triggered data.  By further restricting the range of the fit 
to one can obtain values of $\chi^2$ per degree of freedom much 
closer to unity.  For example, a fit to the $5 < $ \ptt$ < 7$ GeV/c 
for $0.3 < \hat{x}_h < 0.8$ yields a $\chi^2$ per degree of 
freedom of 17.5/5.  However, the extended range of the fit allows 
for a better comparison with the photon triggered data, whose 
poorer statistical precision necessitates fits over a larger range.

\begin{table}[htb]
\caption{Parameters of fits to the \xe\ distributions for \piz\ 
triggers for \xe\ $< 0.8$.}
\begin{ruledtabular} \begin{tabular}{cccc}
\ptt\ [\gevc] & N & $\hat{x}_h$  & $\chi^2$/DOF  \\ \hline
5-7 & 1.25 $\pm$ 0.02 & 0.71 $\pm$ 0.01 & 145/7  \\
7-9 & 1.19 $\pm$ 0.03 & 0.75 $\pm$ 0.01 & 75/7   \\
9-12 & 1.22 $\pm$ 0.06 & 0.76 $\pm$ 0.03 & 18/4  \\
12-15 & 1.33 $\pm$ 0.09 & 0.75 $\pm$ 0.05 & 0.05/2 \\ 
5-15 & 1.24 $\pm$ 0.02 & 0.72 $\pm$ 0.01 & 262/26 \\ 
\end{tabular} \end{ruledtabular} 
\label{tab:xepi0}
\end{table}

For the case of isolated direct photon triggers, Eq.~\ref{eq:mHag} 
will not describe the \xe\ distribution of the away-side jet 
because it is assumed that the trigger particle is also the 
fragment of a jet.  However, for a fixed value of $n$, 
Eq.~\ref{eq:mHag} is a useful quantitative representation of the 
steepness of the \xe\ distribution via the parameter $\hat{x}_{\rm 
h}$; a smaller value of $\hat{x}_h$ corresponds to a steeper 
distribution.  Since the parameters $n$ and $\hat{x}_h$ have 
physical meaning for $\pi^0$-hadron correlations, we keep the same 
value of $n=8.1$ for the isolated-$\gamma$-hadron correlations so 
as to get a quantitative measure of the relative steepness of the 
isolated-$\gamma$-hadron \xe\ distribution compared to the 
$\pi^0$-h distribution.  The results of the fits, shown in 
Table~\ref{tab:xeisodir}, confirm a steeper distribution for photon 
triggers where statistical precision allows.

\begin{table}[htb]
\caption{Parameters of fits to the \xe\ distributions for isolated 
direct photon triggers.  }
\begin{ruledtabular} \begin{tabular}{cccc}
\ptt\ [\gevc] & N & $\hat{x}_h$  & $\chi^2$/DOF  \\ \hline
5-7 & 1.35 $\pm$ 0.14 & 0.60 $\pm$ 0.03 & 2/7  \\
7-9 & 1.40 $\pm$ 0.24 & 0.51 $\pm$ 0.05 & 3/7   \\
9-12 & 0.83 $\pm$ 0.18 & 0.66 $\pm$ 0.10 & 0.5/4  \\
12-15 & 0.75 $\pm$ 0.24 & 0.60 $\pm$ 0.15 & 1.4/2 \\ 
5-15 & 1.11 $\pm$ 0.07 & 0.63 $\pm$ 0.02 & 44/26 \\ 
\end{tabular} \end{ruledtabular} 
\label{tab:xeisodir}
\end{table}

\begin{figure}[htb]
\includegraphics[width=1.0\linewidth]{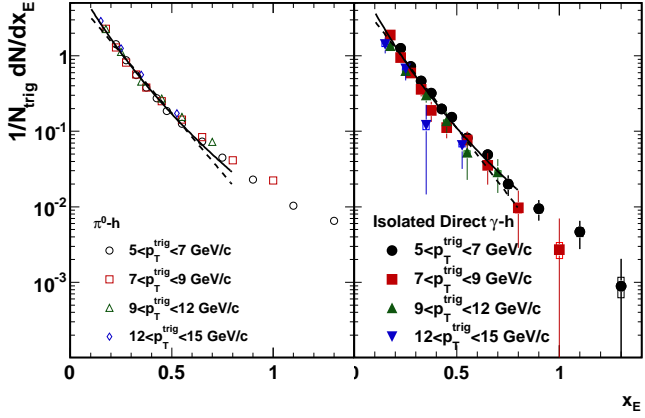} 
\caption[\xe distributions]{\label{fig:xEall} \xe\ distributions 
for \piz\ triggers and isolated direct photon triggers for all 
\ptt\ ranges combined.  $x_E$ is
equivalent to $z_T$ in the collinear limit $cos(\Delta \phi)$ = 1.
The dashed line and solid lines correspond 
to fits to exponential and modified Hagedorn (Eq.~\ref{eq:mHag}) 
functions, respectively.}
\end{figure}

The independence of the $\hat{x}_h$ values for $\pi^0$ as a 
function of $p_T$ suggests that \xe\ scaling should hold for all 
the data combined.  This is shown in the left panel of 
Fig.~\ref{fig:xEall} in which all \piz\ \pt\ selections are fit 
simultaneously.  It is interesting to note that the failure of 
\xe\ scaling in a similar measurement (for lower $p_T $, 2-4 GeV/$c$)
at the CERN-ISR~\cite{CCHK}, led to the concept of parton transverse 
momentum and \kt.

For isolated direct photon production, \xe\ scaling is important 
for a more fundamental reason.  If the \xe\ distribution does indeed 
represent the fragmentation function of the opposite parton, then 
combining all the data (see Fig.~\ref{fig:xEall}) should, apart 
from NLO effects, give a universal distribution which is a 
reasonable representation of the quark fragmentation 
function~\cite{ppg090}.

Within the large errors, the \xe\ scaling appears to hold.  Fits to 
both Eq.~\ref{eq:mHag} and to a simple exponential are shown.  The 
exponential fit ($e^{-b x_E}$)  gives the value $b=8.2\pm 
0.3$, with a $\chi^2$ per degree of freedom of $48/26$, which is in 
excellent agreement with the quark fragmentation function 
parameterized~\cite{ppg029,ppg090} as a simple exponential with 
$b=8.2$ for $0.2<z<1.0$ and inconsistent with the value $b=11.4$ 
for the gluon fragmentation function.
    
\begin{figure}[htb]
\includegraphics[width=1.0\linewidth]{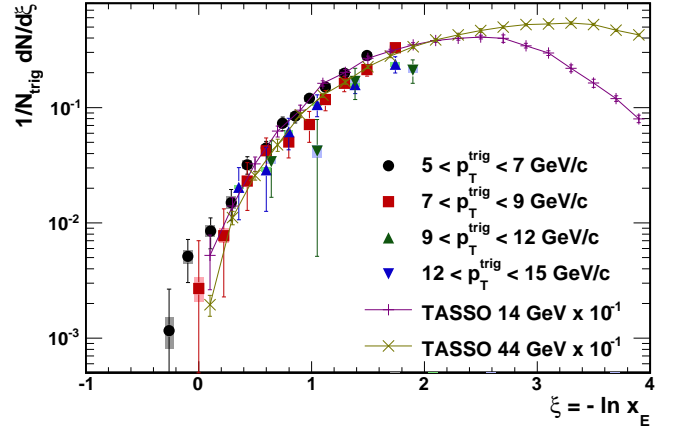} 
\caption[$\xi$ distributions]{\label{fig:alligamxi} $\xi=\ln 
1/x_E$ distributions for isolated direct photon data for all 
\pt\ ranges combined compared to TASSO measurements in $e^+ e^-$ 
collisions at $\sqrt{s}=14$ and 44 GeV.}
\end{figure}

Another, recently more popular way~\cite{BW} to look at the 
fragmentation function is to plot the distribution in the MLLA 
variable~\cite{MLLA} $\xi\equiv \ln 1/z\approx \ln 1/x_E$ 
which is shown in Fig.~\ref{fig:alligamxi}.  The present data 
compare well to the TASSO measurements~\cite{TASSO} in $e^+ + e^-$ 
collisions which have been arbitrarily scaled by a factor of 10 to 
match the PHENIX data, which is reasonably consistent with the 
smaller acceptance of the present measurement.  This again 
indicates consistency with a quark fragmentation function.
         
 \subsection{\pout\ Distributions and  \poutrms\ }
 
\label{sec:pout}

\begin{figure}[htb]
\includegraphics[width=1.0\linewidth]{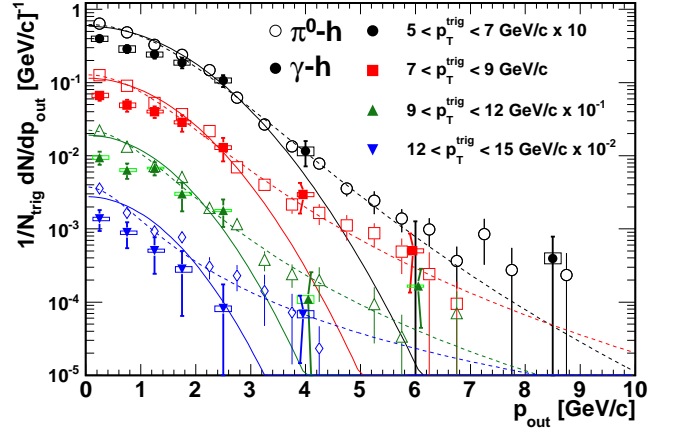}
\caption[]{\pout\ distributions for \piz\ (open symbols) and 
isolated direct photon (filled symbols) triggers.  The \piz\ 
triggered distributions have been fit with both Gaussian (solid 
lines) and Kaplan (dashed lines) functions.  Points with bent error 
bars have been shifted slightly for visibility.  
\label{fig:poutdist}}
\end{figure}

Figure~\ref{fig:poutdist} shows the \pout\ distributions for \piz\ 
and isolated direct photons for the range of $2 < $ \pta\ $ < 10$ 
\gevc.  The \piz\ distributions are fit with Gaussian functions, as 
well as by Kaplan functions.  The Kaplan function is of the form 
$C(1+p_{\rm out}^2/b)^{-n}$, where $C$, $n$ and $b$ are free 
parameters.  This function exhibits the same limiting behavior at 
small values of \pout\ as the Gaussian function and transitions to 
a power-law behavior as \pout\ becomes large.  The tails of the 
distributions, above about 3 \gevc, clearly exhibit a departure 
from the Gaussian fits.  This may signal the transition from a 
regime dominated by multiple soft gluon emission to one dominated 
by radiation of a single hard gluon.  The isolated direct photon 
data also show an excess above the fit, notably for the 
$7 < p_T^{\rm trig} < 9$ \gevc\ range.  For values of \pout\ comparable 
to \ptt, the direct photon data are consistent with zero yield.  
This is to be expected on kinematic grounds as the momentum of 
large angle radiation cannot exceed the jet momentum, which should 
be well-approximated by the photon momentum.

The value of \ktrms\ is determined by measuring \poutrms.  Rather 
than obtaining the values of \poutrms\ directly from the \pout\ 
distributions, fits to the azimuthal correlations are used, 
following the procedure described in \cite{ppg029, ppg089}.  In 
this method the contribution of the underlying event is 
parametrized by a \dphi\ independent contribution and the following 
fit function is used to determine the magnitude of \pout\ from the 
away-side jet width:

\begin{flalign}
\label{eq:poutrms}
& \frac{dN_{\rm real}}{d\Delta\phi} = \nonumber  \frac 1 N \frac{dN_{\rm mix}}{d\Delta\phi}  \cdot \\
& \left(C_0 + C_1\cdot e^{-\Delta \phi^2/2\sigma_{\rm near}^2} + C_2 \cdot \frac{dN_{\rm far}}{d\Delta\phi}
\bigg\arrowvert^{3\pi/2}_{\pi/2}\right)
\end{flalign}

\noindent where

\begin{flalign}
\label{eq:dn_far}
& \frac{dN_{\rm far}}{d\Delta\phi}\bigg\arrowvert^{3\pi/2}_{\pi/2} = \nonumber \\
& \frac{-p_T^{\rm assoc} \cos \Delta \phi}
{\sqrt{2\pi\langle p^2_{\rm out}\rangle}
Erf \left(\sqrt 2 p_T^{\rm assoc}/\sqrt{\langle p^2_{\rm out}\rangle}\right)}
e^{\left( -\frac{|p_T^{\rm assoc}|^2 \sin^2\Delta\phi}{2\langle p^2_{\rm out}\rangle}\right)}.
\end{flalign}

\begin{figure}[htb]
\includegraphics[width=1.0\linewidth]{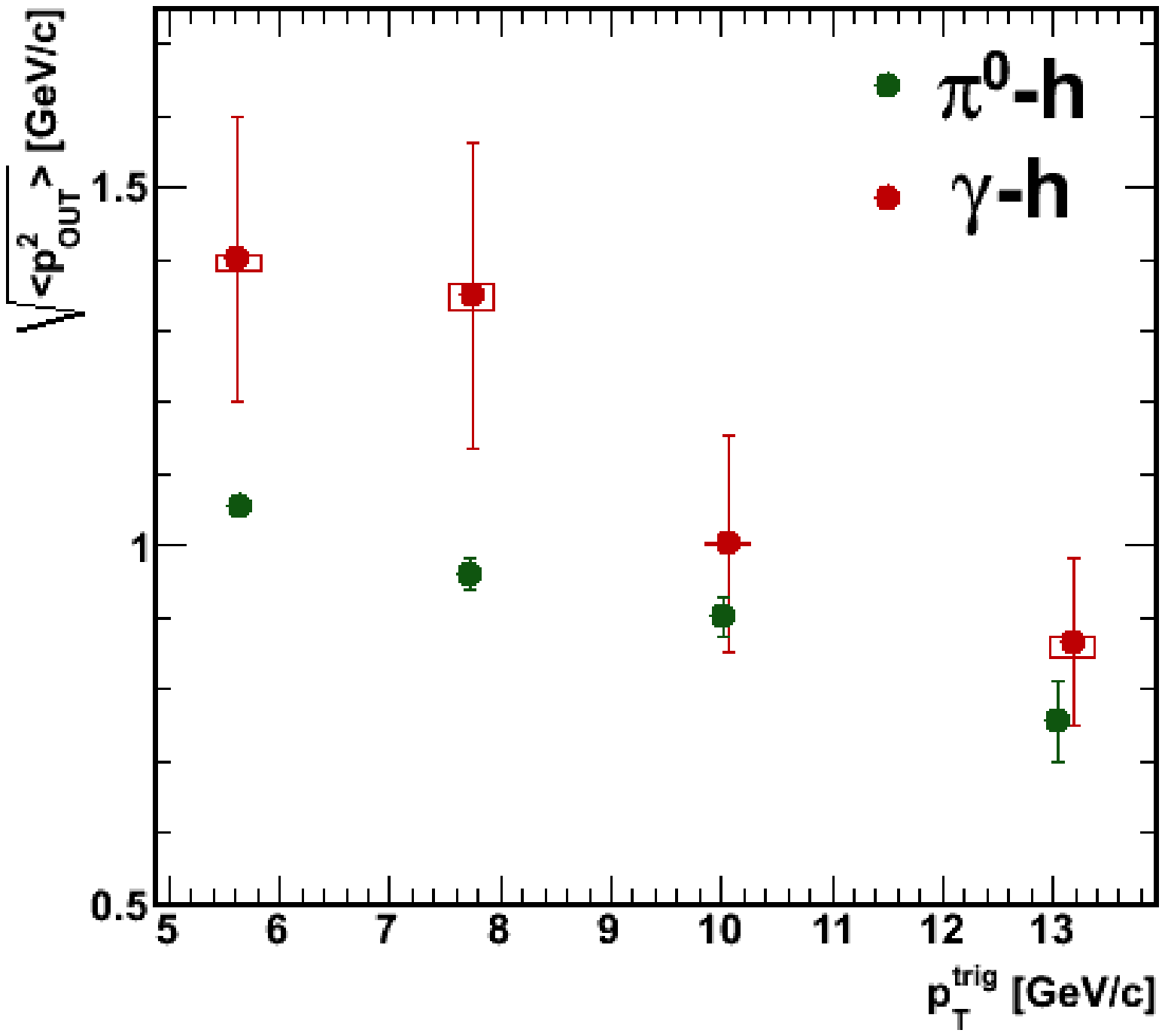}
\caption[]{\poutrms\ for \piz\ and isolated direct photon 
triggers.\label{fig:poutrms_nomc} }
\end{figure}

\noindent The underlying event level $C_0$, the near and away-side 
amplitudes $C_1$ and $C_2$ and \poutrms\ are free parameters.  The 
fits to the \dphi\ distributions are shown in Fig.~\ref{fig:dphi}.  
Using the data for \pta\ $ > 2$ \gevc, we minimize the sensitivity 
to the underlying event whose level is indicated as a dashed line.  
$C_0$ is determined from the \piz\ triggered correlations and 
treated as fixed for the direct photon correlations.  Its value was 
confirmed to be equivalent for both trigger species in the range $1 
< p_T^{\rm assoc} < 2$ \gevc\ where both sets of \dphi\ 
distributions could be reliably fit treating $C_0$ as a free 
parameter.  The \poutrms\ values obtained from the fits are shown 
in Fig.~\ref{fig:poutrms_nomc}.  The width of the \pout\ 
distributions are found to decrease with \ptt.  At the same value 
of \ptt, the isolated direct photon widths are larger than that of 
\piz, which is reasonable given that \piz\ triggers on a larger jet 
momentum.

\section{Discussion and Interpretation}

\subsection{Leading Order + Gaussian \kt\ Smearing Model}

In order to interpret two-particle correlations of final state 
particles in terms of properties of the hard scattered partons, a 
model using the Born level pQCD cross sections and with a Gaussian 
\kt\ smearing was constructed (LO+\kt), similar to the model 
employed in \cite{apanasevich}.  Using this approach, the magnitude 
of the \kt\ effect is varied in the model until the measured values 
of \poutrms\ are reproduced.  At leading order the differential 
cross section for back-to-back hadron production from a $2 
\rightarrow 2$ scattering process ($p_1$+$p_2$ $\rightarrow$ 
$p_3$+$p_4$) is

\begin{flalign}
& \frac{d^5\sigma}{dx_1dx_2d\cos \theta^* dz_3 dz_4} = \nonumber \\ 
& \sum_{\rm a,b,c,d}  f_a(x_1)f_b(x_2)\frac{\pi \alpha_s^2(Q^2)}{2 \hat{s}}\hat{\sigma}_{\rm a,b}(\cos \theta^*) D_c(z_3)
D_d(z_4).  \label{eqn:mcxsec}
\end{flalign}

A Monte-Carlo generator is used to throw flat distributions of 
particle pairs in $x_1, x_2, \theta^*, z_3$ and $z_4$, where $x$ is 
the fraction of the proton's momentum taken by the initial-state 
parton, $\theta^*$ is the polar scattering angle, and $z$ is the 
fraction of the final-state parton taken by a fragment.  For pairs 
which fall in the PHENIX pseudorapidity ($|\eta|<0.35$) interval 
the right-hand side of Equation \ref{eqn:mcxsec} is used to weight 
the contribution of each particle pair.  The expression consists of 
the angular component of 
the LO parton scattering cross section, $\hat{\sigma}$, for each 
parton flavor combination $a+b$ and the non-perturbative parton 
distribution functions (PDF), $f$, and fragmentation functions 
(FF), $D$.  An equal number of all possible permutations of each 
parton flavor are thrown, neglecting charm and heavier quarks.  
The angular part of the parton scattering cross section, 
$\hat{\sigma}_{\rm a,b}(\cos \theta^*)$, contains a numerical 
factor which weights each scattering process appropriately.  
$\hat{s}$ is the Mandelstam variable representing the square of the 
center-of-mass energy at the partonic level and is related to the 
\pp\ center-of-mass energy of $\sqrt{s} = 200$ GeV by $\hat{s} = 
x_1 x_2 s$.  The amplitudes for dijet processes can be found in 
\cite{combridge}.  Direct photon-hadron correlations are also 
considered.  The \gjet\ amplitudes can be found in \cite{fritzsch}.  
In this case, $D(z_4)$ is taken to be a $\delta$ function at $z_4 = 
1$ and the $\alpha_s^{2}$ coupling becomes $\alpha_s \alpha_{\rm EM}$.

The PDFs and FFs are taken from global fit analyses.  The PDFs used 
in this study are the CTEQ6 set \cite{cteq6} evaluated using NLO 
evolution which were obtained from the Durham HEP database 
\cite{durham}.  Several fits to the FFs, KKP, AKK and DSS are 
tested with the Monte Carlo generator \cite{kkp,akk08,dss}.  These 
parametrizations differ in the selection of data sets used in their 
fits.  The KKP set relies solely on $e^+ e^-$ data while the recent 
AKK and DSS fits also employ data from deeply inelastic scattering 
and \pp\ collisions.  The use of these additional data sets enables 
the separation of the quark and anti-quark FFs and provides much 
better constraints on the gluon FFs, although uncertainties for the 
gluon FFs remain large.  Despite the recent progress, the AKK and 
DSS differ quite substantially in a number of observables and 
discrepancies exist between fits and certain data sets.  Newer data 
sets and observables are required to further constrain the FFs.  
Of particular interest for studies of heavy-ion collisions is the 
region of low $z$ where modifications of the FF by the dense QCD 
medium are expected.  Whereas inclusive hadron cross sections are 
most sensitive to relatively large values of $z$ (typically 
0.7-0.8), two-particle correlations provide access to smaller 
values of $z$ using an asymmetric \pt\ selection of the trigger and 
partner.  Direct photons are ideally suited for this purpose due to 
the absence of near-side jet fragmentation effects at LO.  
However, higher order effect such as photon fragmentation and soft 
gluon emission must be constrained.

The \kt\ effect is modeled by introducing a Gaussian distributed 
boost of magnitude $|p_T^{\rm pair}| = \sqrt{2} |k_T|$ randomly 
oriented in the plane transverse to the incoming parton pair, {\it 
i.e.} the beam direction.  The outgoing parton pair, which is 
back-to-back to leading order, hence acquire an acoplanarity 
and a momentum imbalance.  In addition to \kt\ the hadrons also 
acquire some momentum relative to their parent parton direction, 
denoted by \vecjt, from the parton showering process.  The width of 
the \jt\ distribution is taken from measurements of \piz-hadron 
correlations and was determined to be \jtrms\ $= 635$ MeV/$c$ 
\cite{ppg089}.

The Gaussian \kt\ smearing model has the advantage of simplicity 
compared to a more detailed calculation of the underlying jet 
kinematics and subsequent fragmentation.  Although NLO calculations 
are available, modeling the \kt\ smearing would require care to 
avoid double-counting the contribution at NLO with that of soft 
gluon radiation.  A self-consistent approach to doing so would 
require a fully resummed calculation for which generator level 
Monte-Carlo simulations are not presently available.  Although a 
Gaussian \kt\ smearing can be similarly tuned in more sophisticated 
LO models such as PYTHIA, the LO+\kt\ model enables us to test 
various parametrizations of the fragmentation functions as opposed 
to using a model of the hadronization process.

It should be emphasized, however, that the Gaussian \kt\ smearing 
model does not take into account the full parton kinematics and 
hence, requires some care to avoid unphysical scenarios.  The LO 
cross sections are divergent in the forward and/or backward 
directions and the gluon distribution becomes very large at low 
$x$.  In the absence of \kt\ smearing these effects are irrelevant 
for production at midrapidity.  However, for a Gaussian distributed 
\kt\ of fixed width there is a finite probability for a parton to 
be scattered at large angle, solely by virtue of receiving a large 
momentum kick from sampling the tail of the \kt\ distribution.  
Due to the largeness of the low $x$ gluon distribution and the 
cross sections at small angle these soft partons would dominate the 
cross section.  This is clearly an unphysical consequence of the 
\kt\ smearing procedure.  The \kt\ boost is intended to simulate 
gluon emission which should clearly be bounded by the momentum of 
the parton from which it radiates.  This requirement is enforced by 
imposing the that $|k_T| < x \sqrt{s}/2$.  The calculation was 
found to be insensitive to the threshold.

\subsection{Estimating the Magnitude of the \kt\ Effect}

\label{sec:magkt}

\begin{figure}[htb]
\includegraphics[width=1.0\linewidth]{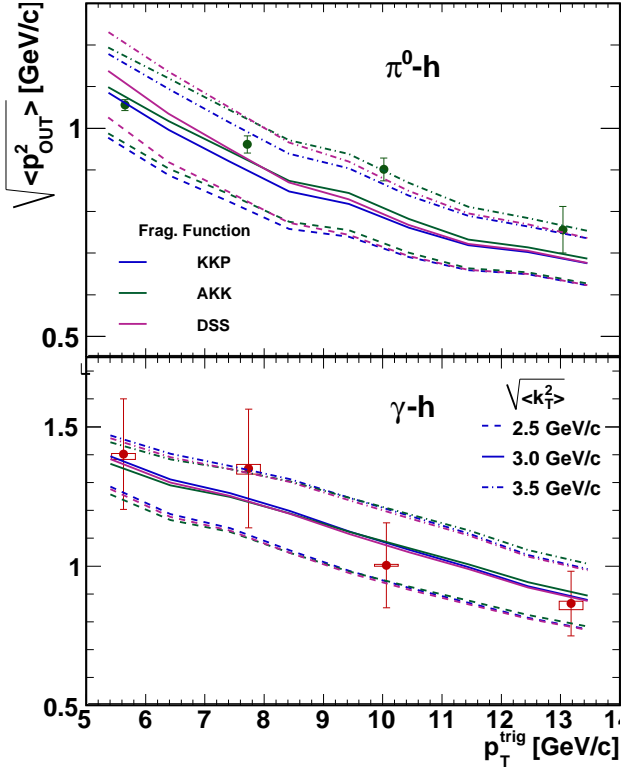}
\caption[\poutrms\ compared to LO+\kt\ calculations]{\poutrms\ 
compared to LO+\kt\ calculations for \piz\ triggers (top) and 
isolated direct photon triggers (bottom) \label{fig:poutrms} }
\end{figure}

In order to determine the best value of \ktrms, \poutrms\ is 
calculated from the model for several different values of \ktrms\ 
and the results are compared to data.  Figure~\ref{fig:poutrms} 
shows \poutrms\ values as well as the LO+\kt\ calculation for 
several values of \ktrms\ and several parametrizations of the FFs.  
The \piz\ triggered data show that although the LO+\kt\ 
qualitatively reproduces the trend of the data, it does not 
perfectly reproduce the \ptt\ dependence of \poutrms.  Additional 
effects not included in the LO+\kt\ smearing model may certainly be 
relevant at this level.  At NLO effects such as the radiation of a 
hard gluon may play a role.  Alternatively, soft gluon radiation 
may not be perfectly described by a Gaussian smearing or may depend 
on the scattering process or the momentum exchange of the hard 
scattering.

\begin{figure}[htb]
\includegraphics[width=1.0\linewidth]{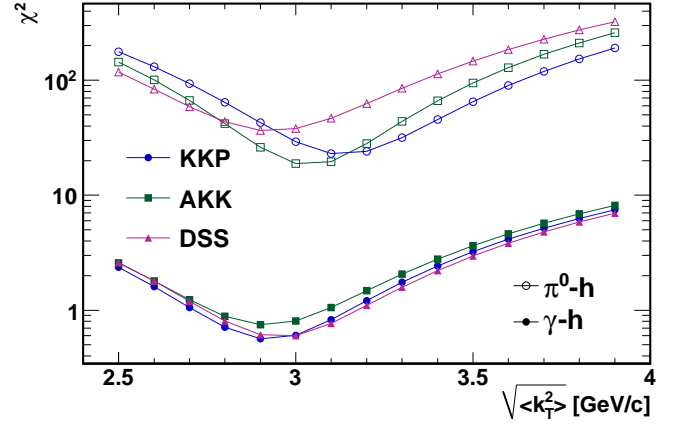}
\caption[$\chi^2$ between \poutrms\ and LO+\kt\ 
calculations]{\label{fig:chi2} $\chi^2$ between the values of 
\poutrms\ obtained in data and from MC calculation for \piz\ and 
isolated direct photon triggers.}
\end{figure}

The $\chi^2$ per degree of freedom between data and model 
calculations are shown in Fig.~\ref{fig:chi2}.  Direct photon 
triggers give a best value of \ktrms\ $\approx 3$ \gevc.  The best 
value for \piz\ triggers is somewhat larger and depends on the 
choice of FF.  This dependence arises from the effect of 
fragmentation on the near-side and the larger fraction of gluon 
jets, for which the parametrizations differ.

\begin{figure}[htb]
\includegraphics[width=1.0\linewidth]{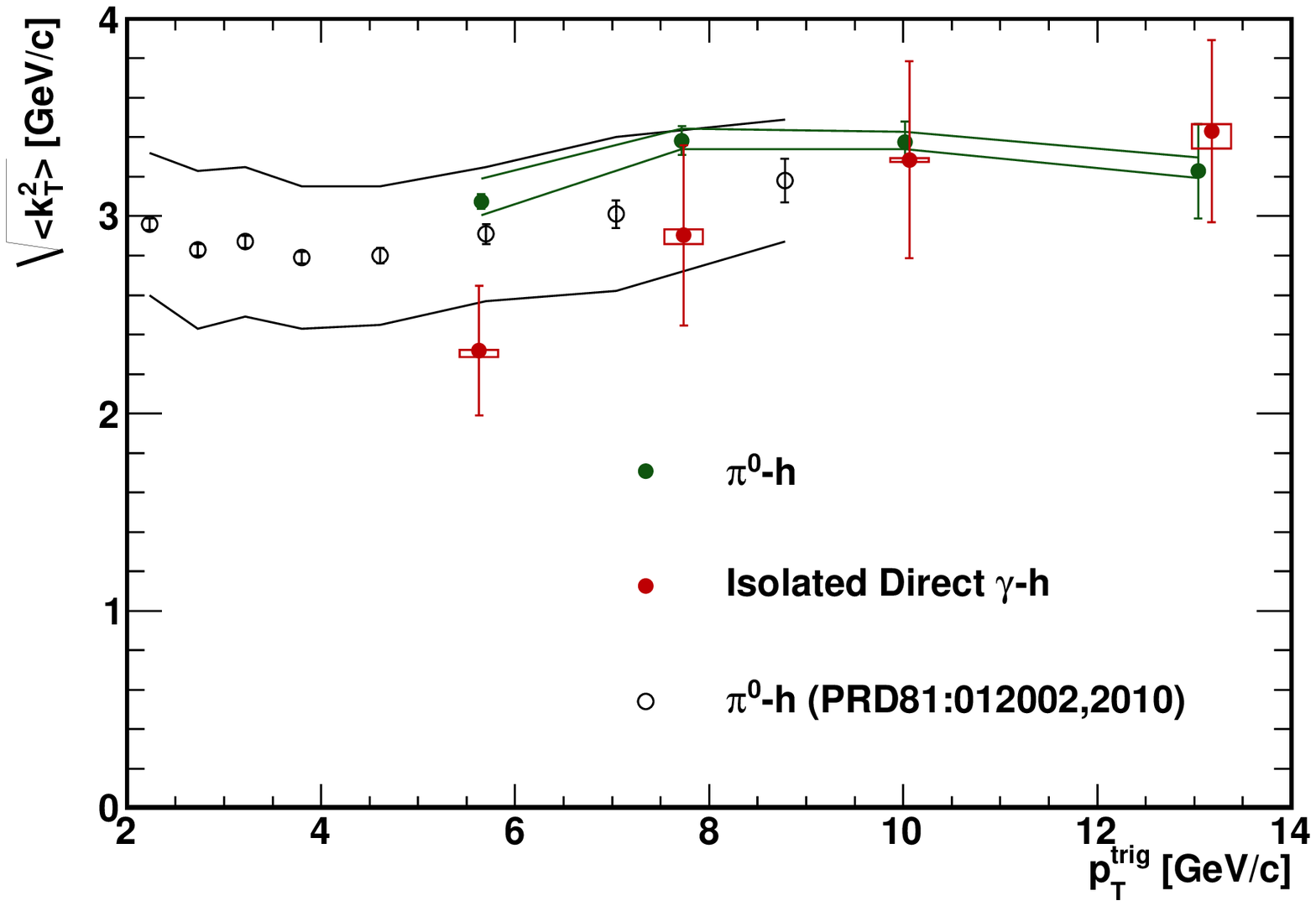}
\caption[]{\ktrms\ as a function of \ptt\ for \piz\ and isolated
direct photon triggers.  Also shown are \pizh\ results using
a different method \cite{ppg029}.  \label{fig:ktvsptt}}
\end{figure}

Figure~\ref{fig:ktvsptt} shows \ktrms\ as function of \ptt\ as 
calculated in the LO+\kt\ model for the best value of the input 
\ktrms\ as determined from Fig.~\ref{fig:chi2}.  The systematic 
error band indicates the dependence on the choice of FF 
parametrization.  The value of \ktrms\ may depend on \ptt\ since 
the trigger requirement may preferentially select events based on 
their \kt.  For both the \piz\ and photon triggered samples, the 
value of \ktrms\ depends on \ptt, and reaches values larger than 
the input, although the effect is only statistically significant 
for the \piz\ triggered data.  The values of \ktrms\ are generally 
larger for \piz\ triggers, but within statistical uncertainties on 
the photon triggered sample, the size of the \kt\ effect is of 
comparable magnitude.

\subsection{Sensitivity of \xe\ Distributions 
to the Fragmentation Functions}

\begin{figure}[tbh]
\includegraphics[width=1.0\linewidth]{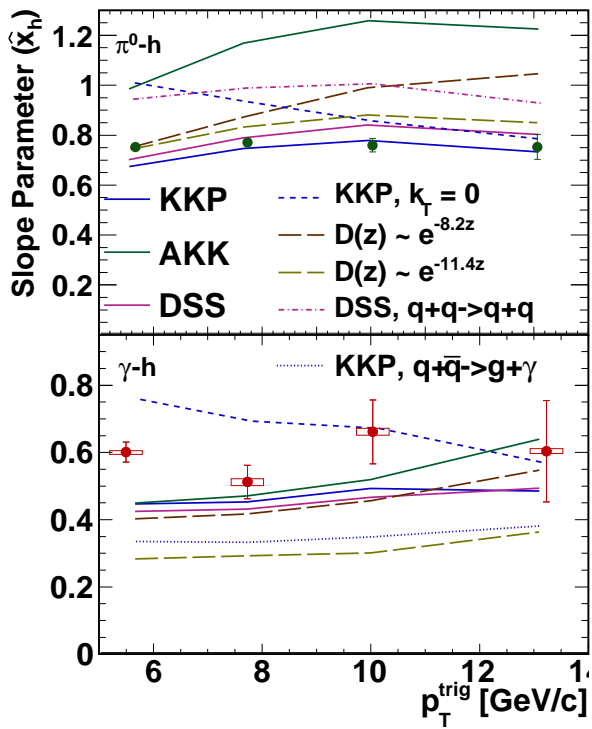}
\caption[]{Slope parameter $\hat{x}_h$ of fits to the \xe\ 
distributions shown in Fig.~\ref{fig:xedists} for \piz\ (top) and 
direct photon (bottom) triggers.  The slopes are compared to 
various calculations using the LO+\kt\ model as discussed in the 
text.}\label{fig:xeslopespi0isodir}
\end{figure}

Using the value of \ktrms\ which best matches the data, the shape 
of the \xe\ distribution can be calculated in the LO+\kt\ model, 
for each set of FFs, and compared to the data.  
Figure~\ref{fig:xeslopespi0isodir} (top panel) shows the slope 
parameter, $\hat{x}_h$, of the power-law fits to the \xe\ 
distributions for \piz\ triggers shown in Fig.~\ref{fig:xedists}.  
Several calculations are shown.  The KKP and DSS fits reproduce the 
shape of the data better than the AKK fit, which shows a much 
harder slope.  The disparity amongst the calculations seems to 
contradict the claim in~\cite{ppg029} that the \xe\ distribution 
for \piz\ triggers is not sensitive to the overall shape of the FF.  
To test this assertion using the LO+\kt\ model, an exponential 
function was used for each flavor of FF.  The slopes of the FFs 
were varied, using $D(z) \propto \exp{(-8.2 z)}$ and $D(z) \propto 
\exp{(-11.4 z)}$, to represent the quark and gluon FFs, 
respectively, as was done in \cite{ppg029}.  Indeed one finds that 
the calculation is not very sensitive to the change in the slope 
parameter.  This exercise does not, however, take into account that 
a change in the shape of the FF for an individual parton flavor may 
change not only the admixture of quark and gluon jets, but also the 
shape of the \pt\ distribution of hard scattered partons, which is 
taken to be fixed in \cite{ppg029}.  Since the PDF for the gluon is 
much different than for the quarks, the sensitivity of the \xe\ 
distribution to the parent parton composition of the \piz\ 
triggered jets can be tested by removing gluon scattering processes 
from the calculation.  To illustrate this, the slope resulting from 
using only the $q+q \rightarrow q+q$ processes with the DSS FFs is 
also shown in Fig.~\ref{fig:xeslopespi0isodir}.  A significantly 
harder slope is obtained, verifying that the \xe\ distribution 
depends on the parton species composition of the triggered jet sample.  
The effect of the \kt\ smearing on the shape of the \xe\ 
distribution was also investigated.  Turning off the smearing 
results in a harder slope for smaller \ptt, which gradually 
disappears as \ptt\ becomes much larger than $\left< k_T \right >$.

Figure~\ref{fig:xeslopespi0isodir} (bottom panel) shows the same 
comparison for isolated photon triggers.  Here the calculation 
depends less strongly on the choice of FF parametrization due to 
the smaller contribution of the annihilation subprocess, $q + 
\bar{q} \rightarrow \gamma + g$, compared to the predominant $g + 
q\rightarrow \gamma + q$.  In contrast to the \piz\ triggered 
sample, the shape of the \xe\ distribution depends rather strongly 
on the shape of the overall FF as demonstrated by varying the slope 
parameter of the exponential parametrization.  The magnitude of 
this effect is reproduced in a gluon jet sample, in which only the 
annihilation process is turned on.  The distribution for the gluon 
jet sample is significantly steeper than the data, verifying that 
the sample is dominated by quark jet fragmentation.  As for the \piz\ 
triggered sample, the Monte-Carlo \xe\ distribution becomes harder 
if the \kt\ smearing is turned off.  Irrespective of the fit, the 
data in the  $5 < p_T^{\rm trig} < 7$ \gevc\ bin are incompatible with
a model without \kt\ smearing, as a significant yield is observed 
at \xe$ > 1$.

\subsection{Charge Asymmetry}

\begin{figure}[htb]
\includegraphics[width=1.0\linewidth]{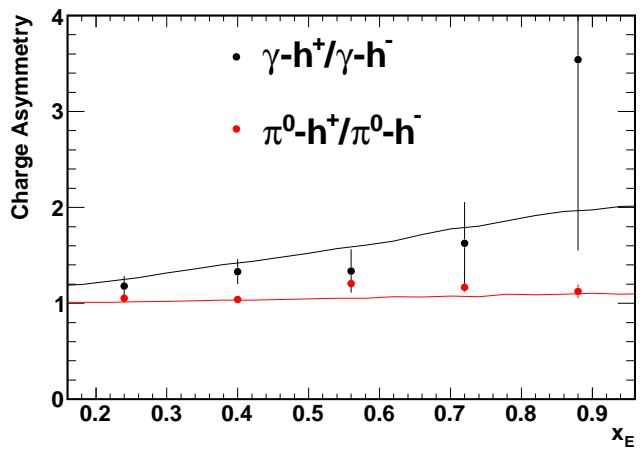}
\caption{Charge asymmetry of hadrons opposite isolated direct 
photon and \piz\ triggers as a function of \xe\ and the 
corresponding calculations in the LO+\kt\ model using the DSS 
FFs.\label{fig:charge_asymm}}
\end{figure}

The dominance of Compton scattering in direct photon production 
implies that the flavor distribution of valence quarks in the 
proton should be reflected in the away-side parton.  Since there 
are twice as many up quarks as down and the amplitude depends on 
the electric charge of the quark, one expects an asymmetry of $8:1$ 
in the number of up quark to down quark recoil jets from the 
Compton scattering of a valence quark and a gluon.  One expects a 
dilution of this factor for several reasons, for example creation 
of charge pairs in the course of the parton shower process, 
corrections from higher order processes such as photon 
fragmentation and a contribution from sea quarks.  Nevertheless, a 
residual charge asymmetry should be apparent in the final state 
hadrons.  Figure~\ref{fig:charge_asymm} shows the ratio of 
positively to negatively charged hadrons $(R\pm)$ on the away-side 
of both \piz\ and isolated direct photon triggers as a function of 
\pta\ along with calculations using the DSS FFs in the LO+\kt\ 
model with the best value of \kt\ as determined in 
Section~\ref{sec:magkt}.  The \piz\ triggered data show an $R\pm$ 
close to unity.  On the other hand, an excess of positive charge is 
evident in the isolated direct photon triggered yields.  This 
supports the claim that the recoil jet is dominated by quark 
fragmentation.

\section{Conclusions}

A detailed understanding of jet fragmentation in \pp\ collisions is 
a prerequisite for studies of possible medium modifications to the 
fragmentation functions by the QCD medium at RHIC.  To this end, 
this study provides baseline measurements of two particle 
correlations from which to quantify these effects.  Direct photon 
triggered correlations were obtained by a combination of 
event-by-event identification, in the form of decay photon tagging 
and isolation cuts, and a subtraction of the residual decay photon 
associated component.  The measurement of near-side hadron 
production associated with inclusive direct photon (i.e., without 
isolation cuts) triggers shows no evidence for a large contribution 
from dijet processes, as might be expected from a large 
fragmentation photon component.  The yield opposite isolated 
direct photons was found to be smaller than for \piz, consistent 
with the expectation of a smaller away-side jet momentum for fixed 
\ptt.  Furthermore, the isolated direct photon associated yields 
were steeper than for \piz\ and demonstrated \xe\ scaling as one 
would expect if the distributions closely resemble the underlying 
charged hadron fragmentation function.  As a function of \pout, the 
away-side yields for \piz\ triggers show a Gaussian-like behavior 
at small values of \pout, whereas a harder, power-law like 
component emerges at large \pout.  The tail component is 
interpreted to be due the emission of a hard gluon.  The isolated 
direct photon triggered distributions also appear to show evidence 
for a hard tail; however the data are consistent with zero yield 
when \pout\ $\geq$ \ptt, corresponding to the kinematic limit for 
LO photon production.

The results were further interpreted at the parton level using a 
simple model of LO pQCD incorporating a 
phenomenologically-motivated Gaussian \kt\ smearing.  The hadron 
yields opposite isolated direct photons are shown to be directly 
sensitive to the fragmentation function of the away-side parton.  
In contrast, hadron triggered jets are sensitive only indirectly, 
due to the contribution of multiple sub-processes in the initial 
state, the relative contribution of which is sensitive to the gluon 
FFs.  Furthermore, the shape of the distributions are shown to be 
compatible with the Compton scattering process $q+g\rightarrow 
q+\gamma$.  The dominance of the Compton scattering process is 
further reinforced by the positive charge asymmetry observed 
opposite isolated direct photon triggers.  Finally, the direct 
photon data are shown to be compatible with a \ktrms\ of similar 
magnitude to that required by the \piz\ triggered data, within 
uncertainties.  Such a large momentum imbalance between the photon 
and the recoil jet is significant for studies of photon tagged jets 
in nuclear collisions in the kinematic regime currently accessible 
at RHIC.


\section*{ACKNOWLEDGMENTS}   

We thank the staff of the Collider-Accelerator and Physics
Departments at Brookhaven National Laboratory and the staff of
the other PHENIX participating institutions for their vital
contributions.  We acknowledge support from the 
Office of Nuclear Physics in the
Office of Science of the Department of Energy,
the National Science Foundation, 
a sponsored research grant from Renaissance Technologies LLC, 
Abilene Christian University Research Council, 
Research Foundation of SUNY, 
and Dean of the College of Arts and Sciences, Vanderbilt University 
(U.S.A),
Ministry of Education, Culture, Sports, Science, and Technology
and the Japan Society for the Promotion of Science (Japan),
Conselho Nacional de Desenvolvimento Cient\'{\i}fico e
Tecnol{\'o}gico and Funda\c c{\~a}o de Amparo {\`a} Pesquisa do
Estado de S{\~a}o Paulo (Brazil),
Natural Science Foundation of China (People's Republic of China),
Ministry of Education, Youth and Sports (Czech Republic),
Centre National de la Recherche Scientifique, Commissariat
{\`a} l'{\'E}nergie Atomique, and Institut National de Physique
Nucl{\'e}aire et de Physique des Particules (France),
Ministry of Industry, Science and Tekhnologies,
Bundesministerium f\"ur Bildung und Forschung, Deutscher
Akademischer Austausch Dienst, and Alexander von Humboldt Stiftung (Germany),
Hungarian National Science Fund, OTKA (Hungary), 
Department of Atomic Energy (India), 
Israel Science Foundation (Israel), 
National Research Foundation (Korea),
Ministry of Education and Science, Russia Academy of Sciences,
Federal Agency of Atomic Energy (Russia),
VR and the Wallenberg Foundation (Sweden), 
the U.S. Civilian Research and Development Foundation for the
Independent States of the Former Soviet Union, 
the US-Hungarian Fulbright Foundation for Educational Exchange,
and the US-Israel Binational Science Foundation.


\begin{thebibliography}{48}
\expandafter\ifx\csname natexlab\endcsname\relax\def\natexlab#1{#1}\fi
\expandafter\ifx\csname bibnamefont\endcsname\relax
  \def\bibnamefont#1{#1}\fi
\expandafter\ifx\csname bibfnamefont\endcsname\relax
  \def\bibfnamefont#1{#1}\fi
\expandafter\ifx\csname citenamefont\endcsname\relax
  \def\citenamefont#1{#1}\fi
\expandafter\ifx\csname url\endcsname\relax
  \def\url#1{\texttt{#1}}\fi
\expandafter\ifx\csname urlprefix\endcsname\relax\def\urlprefix{URL }\fi
\providecommand{\bibinfo}[2]{#2}
\providecommand{\eprint}[2][]{\url{#2}}

\bibitem[{\citenamefont{Ferbel and Molzon}(1984)}]{ferbel}
\bibinfo{author}{\bibfnamefont{T.}~\bibnamefont{Ferbel}} \bibnamefont{and}
  \bibinfo{author}{\bibfnamefont{W.~R.} \bibnamefont{Molzon}},
  \bibinfo{journal}{Rev. Mod. Phys.} \textbf{\bibinfo{volume}{56}},
  \bibinfo{pages}{181} (\bibinfo{year}{1984}).

\bibitem[{\citenamefont{Pumplin et~al.}(2002)}]{cteq6}
\bibinfo{author}{\bibfnamefont{J.}~\bibnamefont{Pumplin}} \bibnamefont{et~al.},
  \bibinfo{journal}{JHEP} \textbf{\bibinfo{volume}{07}}, \bibinfo{pages}{012}
  (\bibinfo{year}{2002}).

\bibitem[{\citenamefont{Aurenche et~al.}()\citenamefont{Aurenche, Baier,
  Fontannaz, Owens, and Werlen}}]{gluondist1}
\bibinfo{author}{\bibfnamefont{P.}~\bibnamefont{Aurenche}},
  \bibinfo{author}{\bibfnamefont{R.}~\bibnamefont{Baier}},
  \bibinfo{author}{\bibfnamefont{M.}~\bibnamefont{Fontannaz}},
  \bibinfo{author}{\bibfnamefont{J.}~\bibnamefont{Owens}}, \bibnamefont{and}
  \bibinfo{author}{\bibfnamefont{M.}~\bibnamefont{Werlen}} (????).

\bibitem[{\citenamefont{Vogelsang and Vogt}(1995)}]{gluondist2}
\bibinfo{author}{\bibfnamefont{W.}~\bibnamefont{Vogelsang}} \bibnamefont{and}
  \bibinfo{author}{\bibfnamefont{A.}~\bibnamefont{Vogt}},
  \bibinfo{journal}{Nucl. Phys.} \textbf{\bibinfo{volume}{B453}},
  \bibinfo{pages}{334} (\bibinfo{year}{1995}).

\bibitem[{\citenamefont{Kniehl et~al.}(2001)\citenamefont{Kniehl, Kramer, and
  Potter}}]{kkp}
\bibinfo{author}{\bibfnamefont{B.~A.} \bibnamefont{Kniehl}},
  \bibinfo{author}{\bibfnamefont{G.}~\bibnamefont{Kramer}}, \bibnamefont{and}
  \bibinfo{author}{\bibfnamefont{B.}~\bibnamefont{Potter}},
  \bibinfo{journal}{Nucl. Phys.} \textbf{\bibinfo{volume}{B597}},
  \bibinfo{pages}{337} (\bibinfo{year}{2001}).

\bibitem[{\citenamefont{Abbott et~al.}(1999)}]{JES1}
\bibinfo{author}{\bibfnamefont{B.}~\bibnamefont{Abbott}} \bibnamefont{et~al.}
  (\bibinfo{collaboration}{D0 Collaboration}), \bibinfo{journal}{Nucl. Instrum.
  Meth.} \textbf{\bibinfo{volume}{A424}}, \bibinfo{pages}{352}
  (\bibinfo{year}{1999}).

\bibitem[{\citenamefont{Ackerstaff et~al.}(1998)}]{phofrag1}
\bibinfo{author}{\bibfnamefont{K.}~\bibnamefont{Ackerstaff}}
  \bibnamefont{et~al.} (\bibinfo{collaboration}{OPAL Collaboration}),
  \bibinfo{journal}{Eur. Phys. J.} \textbf{\bibinfo{volume}{C2}},
  \bibinfo{pages}{39} (\bibinfo{year}{1998}).

\bibitem[{\citenamefont{Bourhis et~al.}(1998)\citenamefont{Bourhis, Fontannaz,
  and Guillet}}]{phofrag2}
\bibinfo{author}{\bibfnamefont{L.}~\bibnamefont{Bourhis}},
  \bibinfo{author}{\bibfnamefont{M.}~\bibnamefont{Fontannaz}},
  \bibnamefont{and} \bibinfo{author}{\bibfnamefont{J.~P.}
  \bibnamefont{Guillet}}, \bibinfo{journal}{Eur. Phys. J.}
  \textbf{\bibinfo{volume}{C2}}, \bibinfo{pages}{529} (\bibinfo{year}{1998}).

\bibitem[{\citenamefont{Alitti et~al.}(1991)}]{isoUA1}
\bibinfo{author}{\bibfnamefont{J.}~\bibnamefont{Alitti}} \bibnamefont{et~al.}
  (\bibinfo{collaboration}{UA2 Collaboration}), \bibinfo{journal}{Phys. Lett.}
  \textbf{\bibinfo{volume}{B263}}, \bibinfo{pages}{544} (\bibinfo{year}{1991}).

\bibitem[{\citenamefont{Abe et~al.}(1994)}]{isoCDF}
\bibinfo{author}{\bibfnamefont{F.}~\bibnamefont{Abe}} \bibnamefont{et~al.}
  (\bibinfo{collaboration}{CDF Collaboration}), \bibinfo{journal}{Phys. Rev.
  Lett.} \textbf{\bibinfo{volume}{73}}, \bibinfo{pages}{2662}
  (\bibinfo{year}{1994}).

\bibitem[{\citenamefont{Abachi et~al.}(1996)\citenamefont{Abachi, Abbott,
  Abolins, Acharya, Adam, Adams, Adams, Ahn, Aihara, Alitti et~al.}}]{isoD0}
\bibinfo{author}{\bibfnamefont{S.}~\bibnamefont{Abachi}},
  \bibinfo{author}{\bibfnamefont{B.}~\bibnamefont{Abbott}},
  \bibinfo{author}{\bibfnamefont{M.}~\bibnamefont{Abolins}},
  \bibinfo{author}{\bibfnamefont{B.~S.} \bibnamefont{Acharya}},
  \bibinfo{author}{\bibfnamefont{I.}~\bibnamefont{Adam}},
  \bibinfo{author}{\bibfnamefont{D.~L.} \bibnamefont{Adams}},
  \bibinfo{author}{\bibfnamefont{M.}~\bibnamefont{Adams}},
  \bibinfo{author}{\bibfnamefont{S.}~\bibnamefont{Ahn}},
  \bibinfo{author}{\bibfnamefont{H.}~\bibnamefont{Aihara}},
  \bibinfo{author}{\bibfnamefont{J.}~\bibnamefont{Alitti}},
  \bibnamefont{et~al.}, \bibinfo{journal}{Phys. Rev. Lett.}
  \textbf{\bibinfo{volume}{77}}, \bibinfo{pages}{5011} (\bibinfo{year}{1996}).

\bibitem[{\citenamefont{Chekanov et~al.}(2004)}]{isoZEUS}
\bibinfo{author}{\bibfnamefont{S.}~\bibnamefont{Chekanov}} \bibnamefont{et~al.}
  (\bibinfo{collaboration}{ZEUS Collaboration}), \bibinfo{journal}{Phys. Lett.}
  \textbf{\bibinfo{volume}{B595}}, \bibinfo{pages}{86} (\bibinfo{year}{2004}).

\bibitem[{\citenamefont{Aaron et~al.}(2008)}]{isoH1}
\bibinfo{author}{\bibfnamefont{F.~D.} \bibnamefont{Aaron}} \bibnamefont{et~al.}
  (\bibinfo{collaboration}{H1 Collaboration}), \bibinfo{journal}{Eur. Phys. J.}
  \textbf{\bibinfo{volume}{C54}}, \bibinfo{pages}{371} (\bibinfo{year}{2008}).

\bibitem[{\citenamefont{Feynman et~al.}(1977)\citenamefont{Feynman, Field, and
  Fox}}]{FFF}
\bibinfo{author}{\bibfnamefont{R.~P.} \bibnamefont{Feynman}},
  \bibinfo{author}{\bibfnamefont{R.~D.} \bibnamefont{Field}}, \bibnamefont{and}
  \bibinfo{author}{\bibfnamefont{G.~C.} \bibnamefont{Fox}},
  \bibinfo{journal}{Nucl. Phys.} \textbf{\bibinfo{volume}{B128}},
  \bibinfo{pages}{1} (\bibinfo{year}{1977}).

\bibitem[{\citenamefont{Altarelli et~al.}(1984)\citenamefont{Altarelli, Ellis,
  Greco, and Martinelli}}]{altarelli}
\bibinfo{author}{\bibfnamefont{G.}~\bibnamefont{Altarelli}},
  \bibinfo{author}{\bibfnamefont{R.~K.} \bibnamefont{Ellis}},
  \bibinfo{author}{\bibfnamefont{M.}~\bibnamefont{Greco}}, \bibnamefont{and}
  \bibinfo{author}{\bibfnamefont{G.}~\bibnamefont{Martinelli}},
  \bibinfo{journal}{Nucl. Phys.} \textbf{\bibinfo{volume}{B246}},
  \bibinfo{pages}{12} (\bibinfo{year}{1984}).

\bibitem[{\citenamefont{Adler et~al.}(2006)}]{ppg029}
\bibinfo{author}{\bibfnamefont{S.~S.} \bibnamefont{Adler}} \bibnamefont{et~al.}
  (\bibinfo{collaboration}{PHENIX Collaboration}), \bibinfo{journal}{Phys. Rev.
  D} \textbf{\bibinfo{volume}{74}}, \bibinfo{pages}{072002}
  (\bibinfo{year}{2006}).

\bibitem[{\citenamefont{Apanasevich et~al.}(1999)}]{apanasevich}
\bibinfo{author}{\bibfnamefont{L.}~\bibnamefont{Apanasevich}}
  \bibnamefont{et~al.}, \bibinfo{journal}{Phys. Rev. D}
  \textbf{\bibinfo{volume}{59}}, \bibinfo{pages}{074007}
  (\bibinfo{year}{1999}).

\bibitem[{\citenamefont{Aurenche et~al.}(2006)\citenamefont{Aurenche, Guillet,
  Pilon, Werlen, and Fontannaz}}]{aurenche}
\bibinfo{author}{\bibfnamefont{P.}~\bibnamefont{Aurenche}},
  \bibinfo{author}{\bibfnamefont{J.~P.}~\bibnamefont{Guillet}},
  \bibinfo{author}{\bibfnamefont{E.}~\bibnamefont{Pilon}},
  \bibinfo{author}{\bibfnamefont{M.}~\bibnamefont{Werlen}}, \bibnamefont{and}
  \bibinfo{author}{\bibfnamefont{M.}~\bibnamefont{Fontannaz}},
  \bibinfo{journal}{Phys. Rev. D} \textbf{\bibinfo{volume}{73}},
  \bibinfo{pages}{094007} (\bibinfo{year}{2006}).

\bibitem[{\citenamefont{Laenen et~al.}(2000)\citenamefont{Laenen, Sterman, and
  Vogelsang}}]{laenen}
\bibinfo{author}{\bibfnamefont{E.}~\bibnamefont{Laenen}},
  \bibinfo{author}{\bibfnamefont{G.}~\bibnamefont{Sterman}}, \bibnamefont{and}
  \bibinfo{author}{\bibfnamefont{W.}~\bibnamefont{Vogelsang}},
  \bibinfo{journal}{Phys. Rev. Lett.} \textbf{\bibinfo{volume}{84}},
  \bibinfo{pages}{4296} (\bibinfo{year}{2000}).

\bibitem[{\citenamefont{Stankus}(2005)}]{stankus}
\bibinfo{author}{\bibfnamefont{P.}~\bibnamefont{Stankus}},
  \bibinfo{journal}{Ann. Rev. Nucl. Part. Sci.} \textbf{\bibinfo{volume}{55}},
  \bibinfo{pages}{517} (\bibinfo{year}{2005}).

\bibitem[{\citenamefont{Adcox et~al.}(2005)}]{whitepaper}
\bibinfo{author}{\bibfnamefont{K.}~\bibnamefont{Adcox}} \bibnamefont{et~al.}
  (\bibinfo{collaboration}{PHENIX Collaboration}), \bibinfo{journal}{Nucl.
  Phys.} \textbf{\bibinfo{volume}{A757}}, \bibinfo{pages}{184}
  (\bibinfo{year}{2005}).

\bibitem[{\citenamefont{Adcox et~al.}(2002)}]{ppg003}
\bibinfo{author}{\bibfnamefont{K.}~\bibnamefont{Adcox}} \bibnamefont{et~al.}
  (\bibinfo{collaboration}{PHENIX Collaboration}), \bibinfo{journal}{Phys. Rev.
  Lett.} \textbf{\bibinfo{volume}{88}}, \bibinfo{pages}{022301}
  (\bibinfo{year}{2002}).

\bibitem[{\citenamefont{Adler et~al.}(2003{\natexlab{a}})}]{ppg014}
\bibinfo{author}{\bibfnamefont{S.~S.} \bibnamefont{Adler}} \bibnamefont{et~al.}
  (\bibinfo{collaboration}{PHENIX Collaboration}), \bibinfo{journal}{Phys. Rev.
  Lett.} \textbf{\bibinfo{volume}{91}}, \bibinfo{pages}{072301}
  (\bibinfo{year}{2003}{\natexlab{a}}).

\bibitem[{\citenamefont{Adler et~al.}(2005)}]{ppg042}
\bibinfo{author}{\bibfnamefont{S.~S.} \bibnamefont{Adler}} \bibnamefont{et~al.}
  (\bibinfo{collaboration}{PHENIX Collaboration}), \bibinfo{journal}{Phys. Rev.
  Lett.} \textbf{\bibinfo{volume}{94}}, \bibinfo{pages}{232301}
  (\bibinfo{year}{2005}).

\bibitem[{\citenamefont{Zakharov}(2004)}]{zakharov}
\bibinfo{author}{\bibfnamefont{B.~G.} \bibnamefont{Zakharov}},
  \bibinfo{journal}{JETP Lett.} \textbf{\bibinfo{volume}{80}},
  \bibinfo{pages}{617} (\bibinfo{year}{2004}).

\bibitem[{\citenamefont{Fries et~al.}(2003)\citenamefont{Fries, Muller, and
  Srivastava}}]{fries}
\bibinfo{author}{\bibfnamefont{R.~J.} \bibnamefont{Fries}},
  \bibinfo{author}{\bibfnamefont{B.}~\bibnamefont{Muller}}, \bibnamefont{and}
  \bibinfo{author}{\bibfnamefont{D.~K.} \bibnamefont{Srivastava}},
  \bibinfo{journal}{Phys. Rev. Lett.} \textbf{\bibinfo{volume}{90}},
  \bibinfo{pages}{132301} (\bibinfo{year}{2003}).

\bibitem[{\citenamefont{Wang et~al.}(1996)\citenamefont{Wang, Huang, and
  Sarcevic}}]{wang}
\bibinfo{author}{\bibfnamefont{X.-N.} \bibnamefont{Wang}},
  \bibinfo{author}{\bibfnamefont{Z.}~\bibnamefont{Huang}}, \bibnamefont{and}
  \bibinfo{author}{\bibfnamefont{I.}~\bibnamefont{Sarcevic}},
  \bibinfo{journal}{Phys. Rev. Lett.} \textbf{\bibinfo{volume}{77}},
  \bibinfo{pages}{231} (\bibinfo{year}{1996}).

\bibitem[{\citenamefont{Adare et~al.}(2009)}]{ppg090}
\bibinfo{author}{\bibfnamefont{A.}~\bibnamefont{Adare}} \bibnamefont{et~al.}
  (\bibinfo{collaboration}{PHENIX Collaboration}), \bibinfo{journal}{Phys. Rev.
  C} \textbf{\bibinfo{volume}{80}}, \bibinfo{pages}{024908}
  (\bibinfo{year}{2009}).

\bibitem[{\citenamefont{Arleo}(2009)}]{mediumFF}
\bibinfo{author}{\bibfnamefont{F.}~\bibnamefont{Arleo}}, \bibinfo{journal}{Eur.
  Phys. J.} \textbf{\bibinfo{volume}{C61}}, \bibinfo{pages}{603}
  (\bibinfo{year}{2009}).

\bibitem[{\citenamefont{Adcox et~al.}(2003{\natexlab{a}})}]{overview}
\bibinfo{author}{\bibfnamefont{K.}~\bibnamefont{Adcox}} \bibnamefont{et~al.}
  (\bibinfo{collaboration}{PHENIX Collaboration}), \bibinfo{journal}{Nucl.
  Instrum. Meth.} \textbf{\bibinfo{volume}{A499}}, \bibinfo{pages}{469}
  (\bibinfo{year}{2003}{\natexlab{a}}).

\bibitem[{\citenamefont{Adcox et~al.}(2003{\natexlab{b}})}]{centralarm}
\bibinfo{author}{\bibfnamefont{K.}~\bibnamefont{Adcox}} \bibnamefont{et~al.}
  (\bibinfo{collaboration}{PHENIX Collaboration}), \bibinfo{journal}{Nucl.
  Instrum. Meth.} \textbf{\bibinfo{volume}{A499}}, \bibinfo{pages}{489}
  (\bibinfo{year}{2003}{\natexlab{b}}).

\bibitem[{\citenamefont{Aphecetche et~al.}(2003)}]{emcal}
\bibinfo{author}{\bibfnamefont{L.}~\bibnamefont{Aphecetche}}
  \bibnamefont{et~al.} (\bibinfo{collaboration}{PHENIX Collaboration}),
  \bibinfo{journal}{Nucl. Instrum. Meth.} \textbf{\bibinfo{volume}{A499}},
  \bibinfo{pages}{521} (\bibinfo{year}{2003}).

\bibitem[{\citenamefont{Adler et~al.}(2007{\natexlab{a}})}]{ppg060}
\bibinfo{author}{\bibfnamefont{S.~S.} \bibnamefont{Adler}} \bibnamefont{et~al.}
  (\bibinfo{collaboration}{PHENIX Collaboration}), \bibinfo{journal}{Phys. Rev.
  Lett.} \textbf{\bibinfo{volume}{98}}, \bibinfo{pages}{012002}
  (\bibinfo{year}{2007}{\natexlab{a}}).

\bibitem[{\citenamefont{Adare et~al.}(2007)}]{ppg063}
\bibinfo{author}{\bibfnamefont{A.}~\bibnamefont{Adare}} \bibnamefont{et~al.}
  (\bibinfo{collaboration}{PHENIX Collaboration}), \bibinfo{journal}{Phys. Rev.
  D} \textbf{\bibinfo{volume}{76}}, \bibinfo{pages}{051106}
  (\bibinfo{year}{2007}).

\bibitem[{\citenamefont{Adler et~al.}(2007{\natexlab{b}})}]{ppg055}
\bibinfo{author}{\bibfnamefont{S.~S.} \bibnamefont{Adler}} \bibnamefont{et~al.}
  (\bibinfo{collaboration}{PHENIX Collaboration}), \bibinfo{journal}{Phys. Rev.
  C} \textbf{\bibinfo{volume}{75}}, \bibinfo{pages}{024909}
  (\bibinfo{year}{2007}{\natexlab{b}}).

\bibitem[{\citenamefont{Adcox et~al.}(2003{\natexlab{c}})}]{tracking}
\bibinfo{author}{\bibfnamefont{K.}~\bibnamefont{Adcox}} \bibnamefont{et~al.}
  (\bibinfo{collaboration}{PHENIX Collaboration}), \bibinfo{journal}{Nucl.
  Instrum. Meth.} \textbf{\bibinfo{volume}{A499}}, \bibinfo{pages}{489}
  (\bibinfo{year}{2003}{\natexlab{c}}).

\bibitem[{\citenamefont{Adler et~al.}(2003{\natexlab{b}})}]{Adler:2003pb}
\bibinfo{author}{\bibfnamefont{S.~S.} \bibnamefont{Adler}} \bibnamefont{et~al.}
  (\bibinfo{collaboration}{PHENIX Collaboration}), \bibinfo{journal}{Phys. Rev.
  Lett.} \textbf{\bibinfo{volume}{91}}, \bibinfo{pages}{241803}
  (\bibinfo{year}{2003}{\natexlab{b}}).

\bibitem[{\citenamefont{Darriulat et~al.}(1976)}]{Darriulat:1976eh}
\bibinfo{author}{\bibfnamefont{P.}~\bibnamefont{Darriulat}}
  \bibnamefont{et~al.}, \bibinfo{journal}{Nucl. Phys.}
  \textbf{\bibinfo{volume}{B107}}, \bibinfo{pages}{429} (\bibinfo{year}{1976}).

\bibitem[{\citenamefont{Della~Negra et~al.}(1977)}]{CCHK}
\bibinfo{author}{\bibfnamefont{M.}~\bibnamefont{Della~Negra}}
  \bibnamefont{et~al.} (\bibinfo{collaboration}{CERN-College de
  France-Heidelberg-Karlsruhe Collaboration}), \bibinfo{journal}{Nucl. Phys.}
  \textbf{\bibinfo{volume}{B127}}, \bibinfo{pages}{1} (\bibinfo{year}{1977}).

\bibitem[{\citenamefont{Borghini and Wiedemann}()}]{BW}
\bibinfo{author}{\bibfnamefont{N.}~\bibnamefont{Borghini}} \bibnamefont{and}
  \bibinfo{author}{\bibfnamefont{U.~A.} \bibnamefont{Wiedemann}},
  \bibinfo{note}{hep-ph/0506218 (2005)}.

\bibitem[{\citenamefont{Dokshitzer et~al.}(1992)\citenamefont{Dokshitzer,
  Khoze, and Troian}}]{MLLA}
\bibinfo{author}{\bibfnamefont{Y.~L.} \bibnamefont{Dokshitzer}},
  \bibinfo{author}{\bibfnamefont{V.~A.} \bibnamefont{Khoze}}, \bibnamefont{and}
  \bibinfo{author}{\bibfnamefont{S.~I.} \bibnamefont{Troian}},
  \bibinfo{journal}{Z. Phys.} \textbf{\bibinfo{volume}{C55}},
  \bibinfo{pages}{107} (\bibinfo{year}{1992}).

\bibitem[{\citenamefont{Braunschweig et~al.}(1990)}]{TASSO}
\bibinfo{author}{\bibfnamefont{W.}~\bibnamefont{Braunschweig}}
  \bibnamefont{et~al.} (\bibinfo{collaboration}{TASSO Collaboration}),
  \bibinfo{journal}{Z. Phys.} \textbf{\bibinfo{volume}{C47}},
  \bibinfo{pages}{187} (\bibinfo{year}{1990}).

\bibitem[{\citenamefont{Adare et~al.}(2010)}]{ppg089}
\bibinfo{author}{\bibfnamefont{A.}~\bibnamefont{Adare}} \bibnamefont{et~al.}
  (\bibinfo{collaboration}{PHENIX Collaboration}), \bibinfo{journal}{Phys. Rev.
  D} \textbf{\bibinfo{volume}{81}}, \bibinfo{pages}{012002}
  (\bibinfo{year}{2010}).

\bibitem[{\citenamefont{Combridge et~al.}(1977)\citenamefont{Combridge,
  Kripfganz, and Ranft}}]{combridge}
\bibinfo{author}{\bibfnamefont{B.~L.} \bibnamefont{Combridge}},
  \bibinfo{author}{\bibfnamefont{J.}~\bibnamefont{Kripfganz}},
  \bibnamefont{and} \bibinfo{author}{\bibfnamefont{J.}~\bibnamefont{Ranft}},
  \bibinfo{journal}{Phys. Lett.} \textbf{\bibinfo{volume}{B70}},
  \bibinfo{pages}{234} (\bibinfo{year}{1977}).

\bibitem[{\citenamefont{Fritzsch and Minkowski}(1977)}]{fritzsch}
\bibinfo{author}{\bibfnamefont{H.}~\bibnamefont{Fritzsch}} \bibnamefont{and}
  \bibinfo{author}{\bibfnamefont{P.}~\bibnamefont{Minkowski}},
  \bibinfo{journal}{Phys. Lett.} \textbf{\bibinfo{volume}{B69}},
  \bibinfo{pages}{316} (\bibinfo{year}{1977}).

\bibitem[{dur()}]{durham}
\bibinfo{note}{Http://durpdg.dur.ac.uk/hepdata/pdf3.html}.

\bibitem[{\citenamefont{Albino et~al.}(2008)\citenamefont{Albino, Kniehl, and
  Kramer}}]{akk08}
\bibinfo{author}{\bibfnamefont{S.}~\bibnamefont{Albino}},
  \bibinfo{author}{\bibfnamefont{B.~A.} \bibnamefont{Kniehl}},
  \bibnamefont{and} \bibinfo{author}{\bibfnamefont{G.}~\bibnamefont{Kramer}},
  \bibinfo{journal}{Nucl. Phys.} \textbf{\bibinfo{volume}{B803}},
  \bibinfo{pages}{42} (\bibinfo{year}{2008}).

\bibitem[{\citenamefont{de~Florian et~al.}(2007)\citenamefont{de~Florian,
  Sassot, and Stratmann}}]{dss}
\bibinfo{author}{\bibfnamefont{D.}~\bibnamefont{de~Florian}},
  \bibinfo{author}{\bibfnamefont{R.}~\bibnamefont{Sassot}}, \bibnamefont{and}
  \bibinfo{author}{\bibfnamefont{M.}~\bibnamefont{Stratmann}},
  \bibinfo{journal}{Phys. Rev. D} \textbf{\bibinfo{volume}{75}},
  \bibinfo{pages}{114010} (\bibinfo{year}{2007}).

\end{thebibliography}

\end{document}